\documentclass{aa}
\usepackage[usenames]{color}
\usepackage{mathrsfs}
\usepackage{txfonts}
\usepackage{xcolor}
\usepackage{graphicx}
\usepackage{natbib}
\usepackage{enumitem}
\usepackage[utf8]{inputenc} 
\usepackage[T1]{fontenc}    

\bibpunct{(}{)}{;}{a}{}{,} 

\newcommand{\Mm}{{\mathrm{\, Mm}}}
\newcommand{\kms}{{\mathrm{\, km \,s^{-1}}}}

\newcommand{\nstr}{{{N_\mathrm{s}}}}
\newcommand{\rstr}{{{r_\mathrm{s}}}}
\newcommand{\rhostr}{{{\rho_\mathrm{s}}}}
\newcommand{\pstr}{{{P_\mathrm{s}}}}
\newcommand{\plow}{P_\mathrm{L}}
\newcommand{\phigh}{P_\mathrm{H}}

\newcommand{\kzva}{k_\mathrm{z} \, v_\mathrm{A}}

\DeclareGraphicsExtensions{.pdf,.png,.jpg}

\begin{document}
\title{Fundamental Transverse Vibrations of the Active Region Solar Corona}

\date{Received <date> /
Accepted <date>}

\author{M. Luna \inst{1,2}
          \and
          R. Oliver \inst{3,4}
          \and
          P. Antolin \inst{5}
          \and
          I. Arregui \inst{1,2}
          }

   \institute{Instituto de Astrof\'{\i}sica de Canarias, E-38205 La Laguna, Tenerife, Spain\\
              \email{mluna@iac.es}
         \and
             Departamento de Astrof\'{\i}sica, Universidad de La Laguna, E-38206 La Laguna, Tenerife, Spain
             \and
             Departament F{\'i}sica, Universitat de les Illes Balears, E-07122 Palma de Mallorca, Spain
             \and
             Institute of Applied Computing \& Community Code (IAC$^3$), UIB, Spain
             \and
             University of St Andrews, St Andrews, KY16 9AJ, United Kingdom
             }

\titlerunning{Vibrations of the Solar Corona}
\authorrunning{Luna, Oliver, Antolin \& Arregui}

\abstract {Some high-resolution observations have revealed that the active-region solar corona is filled with myriads of thin strands even in apparently uniform regions with no resolved loops. This fine structure can host collective oscillations involving a large portion of the corona due to the coupling of the motions of the neighbouring strands.} {We study these vibrations and the possible observational effects.} {Here we theoretically investigate the collective oscillations inherent to the fine structure of the corona. We have called them fundamental vibrations because they cannot exist in a uniform medium. We use the T-matrix technique to find the normal modes of random arrangements of parallel strands. We consider an increasing number of tubes to understand the vibrations of a huge number of tubes of a large portion of the corona. We additionally generate synthetic time-distance Doppler and line broadening diagrams of the vibrations of a coronal region to compare with observations.}{We have found that the fundamental vibrations are in the form of clusters of tubes where not all the tubes participate in the collective mode. The periods are distributed over a wide band of values. The width of the band increases with the number of strands but rapidly reaches an approximately constant value. We have found an analytic approximate expression for the minimum and maximum periods of the band. The frequency band associated with the fine structure of the corona depends on the minimum separation between strands. We have found that the coupling between the strands is of large extent and the motion of one strand is influenced by the motions of distant tubes. The synthetic Dopplergrams and line-broadening maps show signatures of collective vibrations, not present in the case of purely random individual kink vibrations.} {We conclude that the fundamental vibrations of the corona can contribute to the energy budget of the corona and they may have an observational signature.}


\maketitle

\section{Introduction and motivation}\label{sec:intro}
Early X-ray and extreme ultra-violet (EUV) observations by rocket bourne instruments \citep{vaiana1968,vanspeybroeck1970} and onboard Skylab \citep{tousey1973,vaiana1973} already revealed that the active region (AR) corona is formed by myriads of coronal loops. Because of the high anisotropy of the solar corona, and in particular that of transport coefficients whose parallel components to the field are much larger than their perpendicular components, it is thought that field-aligned structures should evolve largely independently on a given timescale \citep{klimchuk2015}. This suggests the existence of coronal strands, that is, elemental flux tubes, as the building blocks of coronal loops and the entire solar corona. This concept has received support either directly from some recent high resolution observations, or indirectly from the combination of observations and modelling. Indeed, several works using SDO/AIA \citep{lemen2011}, Hinode/EIS \citep{tsuneta2008}, Hi-C \citep{cirtain2013} and IRIS \citep{pontieu2014} have found strand sizes of few hundreds of km \citep[see, e.g.][]{peter2013,brooks2012,brooks2013,brooks2016}. \citet{antolin2012} observed strand-like structures of a few hundred km in observations of coronal rain with the SST/CRISP \citep{scharmer2008} instrument. \citet{antolin2015} have shown that the distribution of rain strands peaks at about 200~km and falls abruptly at the resolution limit, suggesting that even lower sizes are present \cite[see also][]{scullion2014}. This tip of the iceberg distribution is further supported by observations of flare-driven rain at even higher resolution with the NST \citep{Jing_2016NatSR...624319J}, in which the distribution peaks at 120~km with again a sharp cut-off at smaller sizes.   
Some authors claim that strands have radii as small as few tens of kilometres \citep{peter2013}. Hence, it is possible that coronal loops and, more generally the corona, are structured in the form of myriads of thin flux tubes called strands.

The volumetric filling factor, $f = V_s/V$, measures the portion of a coronal region volume, $V$, filled with strands with a total volume $V_s$. The filling factor can be estimated using spectroscopic observations. \citet{warren2008} found volumetric filling factors of $10 \%$ in AR loops. The authors also found a very narrow band of temperatures and densities. \citet{landi2009} find a filling factor of $30\%$ in cooling loops. \citet{young2011} find filling factors between $3 - 30\%$ but the most reliable pixels show a filling factor between $10 - 20\%$. Elucidating how the solar corona is structured is crucial to understand the mechanisms that heat this upper part of the Sun's atmosphere \citep{klimchuk2015}. 
 
The solar atmosphere is very dynamic and it has been observed that energetic disturbances such as flares, coronal mass ejections or low coronal eruptions can excite transverse oscillations in magnetic loops \citep[see, e.g.,][]{nakariakov1999,aschwanden1999,zimovets2015}. At smaller amplitudes these waves are observed ubiquitously \citep{tomczyk2007,tomczyk2009,anfinogentov2015,morton2019}. Based on magnetohydrodynamic (MHD) wave theory \citep[see review by][]{nakariakov2005}, transverse loop oscillations have been interpreted in terms of (fast) kink MHD waves. 

The ubiquitous existence of kink waves in the solar atmosphere has further ignited the debate of the strand concept as fundamental constituents of coronal loops. While preferential field aligned transport of mass and energy may be enough to validate the concept of strands, \citet{magyar2016} have shown that the effect of kink waves on previously existing strand-like substructure in a loop is to destroy them on a timescale of a few wave periods. This is due to the mixing produced by the Kelvin-Helmholtz instability (KHI) generated by the velocity shear at the edges of the loop. On the other hand, \citet{antolin2014} have shown that even monolithic loops may show a strand-like structure in EUV intensity images due to the  transverse wave induced Kelvin-Helmholtz (TWIKH) rolls generated by the waves. In any case, as argued by \citet{VanDoorsselaere_2018AA...620A..65V}, the very concept of an isolated and independently evolving structure, be it a strand or a monolithic loop is compromised by transverse MHD waves since their fast nature makes them collective and can, therefore, couple the (thermo-)dynamic evolution of various such structures.

Aside from the debate on the structure of individual coronals loops, in this work we hypothesise that the AR corona is formed by myriads of thin flux tubes or strands even in an apparently uniform region with no resolved coronal loops (i.e., the diffuse corona). Following this hypothesis, we study the collective oscillations inherent to the multi-stranded AR corona. The motions of each individual strand are not isolated from the motions of their neighbours \citep[see,][]{luna2006,luna2008,luna2009,luna2010}. If a strand is perturbed, the surrounding plasma also moves to produce the motion of the nearby tubes. In this sense, the motions of the strands are coupled and periodic motions are in the form of collective oscillations. Due to this coupling, large regions of the corona may oscillate collectively in phase. The extension of these collective oscillations, the way the strands oscillate or the period of the oscillations will be related to the fine structure of the corona. Hence, these collective oscillations are inherent to the multi-stranded fine structure of the corona.

Past research on collective oscillations in mult-stranded systems has shown that the interaction between neighbouring tubes can modify the properties of their transverse oscillations with respect to the properties of the kink mode of an isolated tube. Several works have considered transverse loop oscillations in slab geometry \citep{diaz2005,diaz2006,luna2006,arregui2007,arregui2008} or in cylindrical geometry \citep{luna2008,van-doorsselaere2008,ofman2009,luna2009,luna2010,esmaeili2016} with several configurations of two or more tubes. In general, these studies show that new collective oscillatory modes with complex spatial structure appear in multiple tube configurations. As shown by \cite{luna2010}, none of the modes produces global kink motion with all the strands moving in phase in the same direction. When numerically simulating the temporal evolution of a system of 10 strands, \cite{luna2010} find that the system initially oscillates in phase in the direction of the initial disturbance. After some time, this organised motion disappears and the complexity of the motions of the strands increases. The frequencies of the individual modes split into many new collective frequencies. This splitting depends on the distance between the tubes and  their properties. The coupling between tubes and consequently the frequency splitting is maximum when the densities of the tubes are identical \citep{luna2009}. 

Most of these studies were applied to coronal loop oscillations but their results can also be applied to bundles of strands in a region of the AR corona. Our hypothesis defines these vibrations as fundamental because these motions are associated to the fine structure of the corona. Several mechanisms, such as flares, coronal eruptions, or in general any impulsive disturbance that can deposit energy in the corona, can trigger the vibrations of coronal strands. Additionally, the continuous buffeting of waves from the bottom layers of the solar corona can also excite such collective oscillations. These collective vibrations are probably difficult to measure due to the combination of an optically thin plasma and the complex movements of those structures. In the same line-of-sight (LOS), the strands move in many directions producing probably a small net or integrated velocity and subsequently a small Doppler signal. \cite{mcintosh2012} found that the corona has a ``dark'' wave energy content in the form of non-thermal line broadening that is consistent with periodic motions. However, the periodic motions considered in that work did not take into account the coupled nature of fundamental vibrations. One of the main goals of this paper is precisely to show that the normal modes that we describe here can also contribute to this hidden wave energy content. There are other wave processes that can contribute to the dark energy, as the resonant absorption (or mode coupling), as shown by \citet{moortel2012,antolin2017}. These authors have shown that only 10\% of the kinetic energy of the waves could be observed due to the combination of resonant absorption (or mode coupling process) and LOS superposition.

The layout of the paper is as follows. In Section \ref{sec:model} we describe the model we use to compute these collective normal modes. In Section \ref{sec:clustering} we study the normal modes of a large number of strands. We study how the modes depend on the size of the strand system in Section \ref{sec:depencewithsystemsize} and an analytical approach for the period splitting is shown. In Section \ref{sec:syntheticimages} we generate synthetic images of the vibrations of the corona. Finally, in Section \ref{sec:conclusions} conclusions of this work are drawn.

\section{Theoretical model}\label{sec:model}
The objective of this work is to explore the vibrations of the AR corona associated with its multi-stranded fine structure. The solar corona is composed of myriads of strands and the collective oscillations can involve the motion of a huge, virtually infinite, number of tubes. A natural attempt can be to consider an infinite and periodic structure of strands. This oversimplification can help to find the normal modes due to the symmetry of the problem. Due to the spatial periodicity, the normal modes will involve the motion of all the strands of the modelled corona. This periodic arrangement is very unrealistic. A random disposition of the strands would be more realistic, however, to consider a virtually infinite number of strands with random positions is not feasible. For this reason, we consider a finite number of strands arranged randomly in a region of the corona.  We have defined a strategy to study the normal modes of a system with progressively increasing size.
In this way, we can understand how the normal modes in a system with a large number of strands are established. A pattern emerges that can be extrapolated to a virtually infinite system of strands.
Additionally, considering a finite region of the corona the oscillations of the strands inside the region are coupled with oscillations outside the region. Thus, in principle, it is not possible to isolate the motions of the strands of such region from the tubes of its environment. For these reasons, the system should be finite but large enough so that the motions in a delimited region are representative of the collective oscillations in the AR corona.


This work is a proof of concept and we keep the system as simple as possible. Our equilibrium configuration consists of $\nstr$ identical straight and parallel magnetic cylindrical strands of length $L$ embedded in a uniform coronal plasma. Each coronal strand has a uniform density, $\rhostr$, along the tube (gravity is neglected) with the loop feet tied in the photosphere. However, in Section \ref{sec:non-identicalstrands} we study the influence of strand lengths and densities differences on the collective modes. The $z$-axis points in the direction of the strand axes. The equilibrium magnetic field is straight and uniform along the $z$-direction, ${\bf B}=B \, \hat{{\bf e}}_{z} $. All the strands have an identical radius, $\rstr$. The volumetric filling factor, $f = V_s/V=\nstr \, \rstr^{2}/R^{2}$, measures the portion of a coronal region volume, $V$, filled with strands with a total volume $V_s$. The coronal region of volume $V$ could have any shape but we have assumed a cylindrical volume with radius $R$ for simplicity. This cylinder is physically meaningless in this work and it just delimits the region where the strands are placed. We also assume no boundary layer between the strand and the external corona (i.e. only a Heavyside function). This therefore eliminates the resonant absorption mechanism in our model.

\begin{figure}[!ht]
\begin{center}
\centering\includegraphics[width=0.4\textwidth]{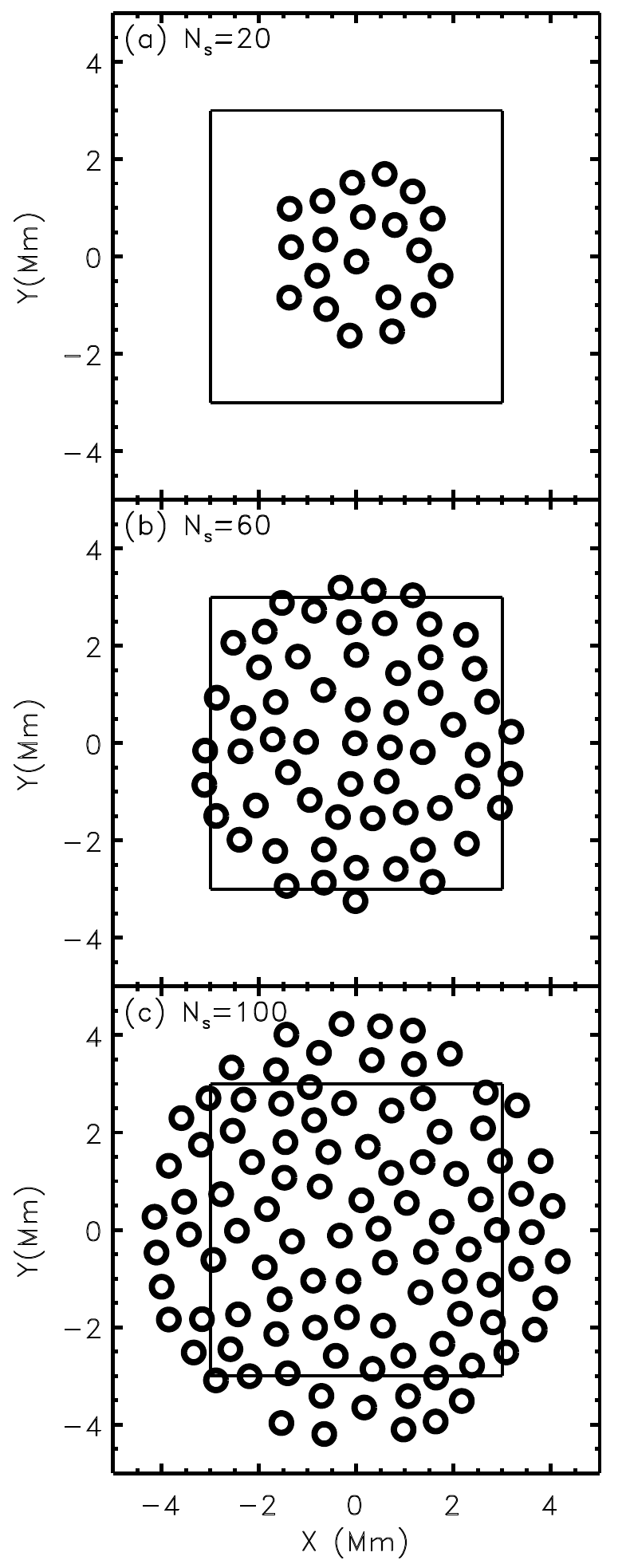}
\caption{Sketch of the cross-section of a multi-stranded AR model, which consists of (a) $\nstr=20$, (b) $\nstr=60$ and (c) $\nstr=100$  homogeneous strands of densities $\rhostr$ and radii $\rstr$. In the figure, the strands are represented by their external surface of radius $\rstr$ (thick solid lines). The external medium to the strands consists of coronal material with density $\rho_\mathrm{c}$. }\label{fig:identical-strands-spatial-distribution}\label{fig:model}
\end{center}
\end{figure}

In our study we consider different distributions with $\nstr=$1, 5, 10, 20, 30, 40, 50, 60 and 100 parallel strands. We use the T-matrix technique to find the normal modes of the system as explained in Section \ref{sec:clustering}. Considering systems larger than 100 strands is difficult due to numerical reasons. In Figure \ref{fig:model} we have plotted the strand distribution for 20, 60 and 100 strands. In addition, the region within the square delimits an area of $6\times 6 \Mm^2$ representative of a small portion of an AR corona. In principle the size of this region is arbitrary but we have selected this size in order to have a sufficiently large number of tubes inside. Henceforth, this area is our region of interest or ROI for abbreviation. From the figure, we see that only the 100 strands configuration covers the ROI. We will see that the vibrations in the square area are not fully described by the normal modes of the 100 strands system. However, with this limited system of 100 strands, we can understand the complexity of the oscillations of the strands in the ROI.

The strands have a length $L=100\Mm$ and radius $\rstr=0.2\Mm$ that are typical values in agreement with observations (see Sec. \ref{sec:intro}). Each individual strand, labeled as $j$, is placed in the $xy$-plane in a  position $\vec{r}_{j}=x_{j} \, \vec{e}_{x}+y_{j} \, \vec{e}_{y}$. These positions in the $xy$-plane are randomly generated inside the circular area of radius $R$. In order to keep a constant filling factor, $R= \rstr \,\sqrt{\nstr/f}$. These positions should fulfil the condition $|\vec{r}_{j}|\le R - \rstr$ in order to have all the strands inside the circular area with radius $R$. In addition, the minimum distance between any two strands $i$ and $j$ is $d_{i j} \ge d_\mathrm{min}=\mathcal{S} \, 2 \, \rstr $. With the arbitrary parameter $\mathcal{S}$, we control the minimum distance between the strands. With $\mathcal{S}=1$ the surfaces of the strands can be in contact.  We have chosen $\mathcal{S}=1.692$ in order to have a more or less uniform distribution of strands. In this situation  $d_\mathrm{min}=0.68 \Mm$. We additionally consider a typical filling factor $f=0.2$ in agreement with the observed values (see Sec. \ref{sec:intro}). For the particular cases shown in Figure \ref{fig:model}, the strands cover a circular surface of radius $R=2.00, 3.46$ and $4.47 \Mm$ respectively.  We assume a strand density $\rhostr=3.5 \, \rho_\mathrm{c}$ in terms of the background coronal density, being $\rho_\mathrm{c}=2 \times 10^{-13} \, \mathrm{kg \, m^{-3}}$, and a magnetic field intensity $B=5$ Gauss. With these parameters, the individual kink period of each strand is $\pstr$= 5 min. This period is a typical value for loop oscillations but also for the transverse waves of the solar corona \citep[e.g.,][]{tomczyk2007}.

\section{Collective normal modes: clustering}\label{sec:clustering}
We first show how the normal modes establish in a large system of strands. We consider the example with $\nstr=60$ strands shown in Figure \ref{fig:model}(b).
\begin{figure}[!ht]
\centering\includegraphics[width=0.5\textwidth]{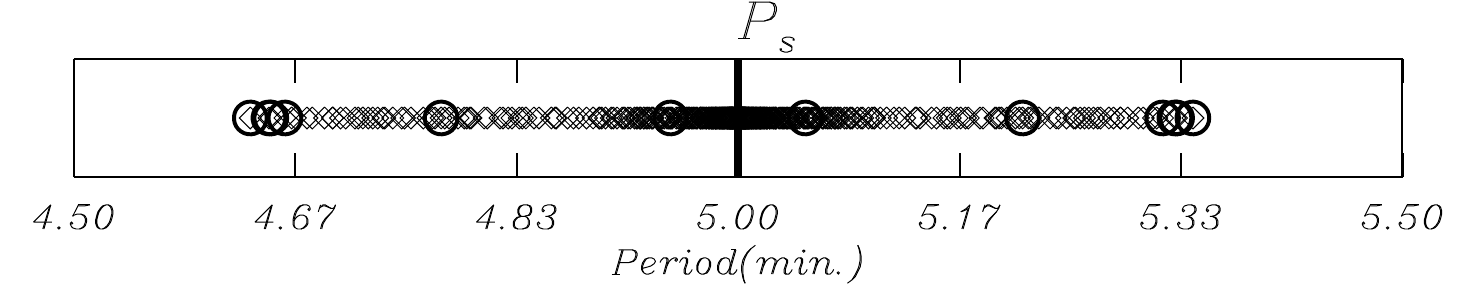}
\caption{Frequency distribution of the collective normal modes associated with the $\nstr=60$ case. We clearly see a distribution of periods around $\pstr=5 \min$. The circles mark the periods of the modes whose spatial structure is displayed in Fig. \ref{fig:multipanel-modes-60}.\label{fig:periods60}}
\end{figure}
The normal modes of the system depend on the number of strands and their spatial distribution \citep{luna2009,luna2010}. We have found the normal modes of all the strands system using the T-matrix technique by \citet{luna2009}. In general, in the strand systems, the individual period $\pstr$ splits in a virtually infinite number of frequencies. The collective frequencies are distributed at both sides of the individual strand frequency (see Fig. \ref{fig:periods60}). According to \citet{luna2010} the normal modes can be classified according to their frequency: modes with frequencies below the central frequency, i.e. $2\, \pi / \pstr$, are called Low modes (right-hand side of the $\pstr$ line in Fig. \ref{fig:periods60}). Mid modes are those with periods similar to $2\, \pi / \pstr$, and finally, the solutions with frequencies above the central frequency are called High modes (at the left-hand side of the $\pstr$ line in Fig. \ref{fig:periods60}). In the Low modes the fluid between the tubes follows the motion of the strands, resulting in a small perturbation of the magnetic pressure. In contrast, in the High modes, the fluid between tubes is compressed or rarefied with important perturbations of the pressure. Mid modes consist of more complex motions of the strands. The distinction between Low and High is  straightforward because both have frequencies lower or higher than the individual kink mode frequency respectively. However, there is  no clear transition between Low and Mid modes or Mid and High modes. In terms of periods, the collective oscillations are between the maximum period, $\plow$, associated to the Low mode with the minimum frequency and the minimum period, $\phigh$, associated to the High mode with the largest frequency. Note that the notation High and Low refers to the frequency and $\plow$ and $\phigh$ correspond to periods with the highest and lowest periods values respectively so that the subscripts refer to the mode frequency, from which the period is derived.  The period band has a width $\Delta P= \plow-\phigh$. In the case of 60 strands shown in Figure \ref{fig:periods60}, the period band ranges from $\phigh=4.63\min$ to $\plow=5.34\min$ with a bandwidth $\Delta P= 0.71\min$. In terms of frequencies, the modes range from 3.12 to 3.6 mHz that is a frequency split of 0.48 mHz.

\begin{figure*}[!ht]
\centering\includegraphics[width=1\textwidth]{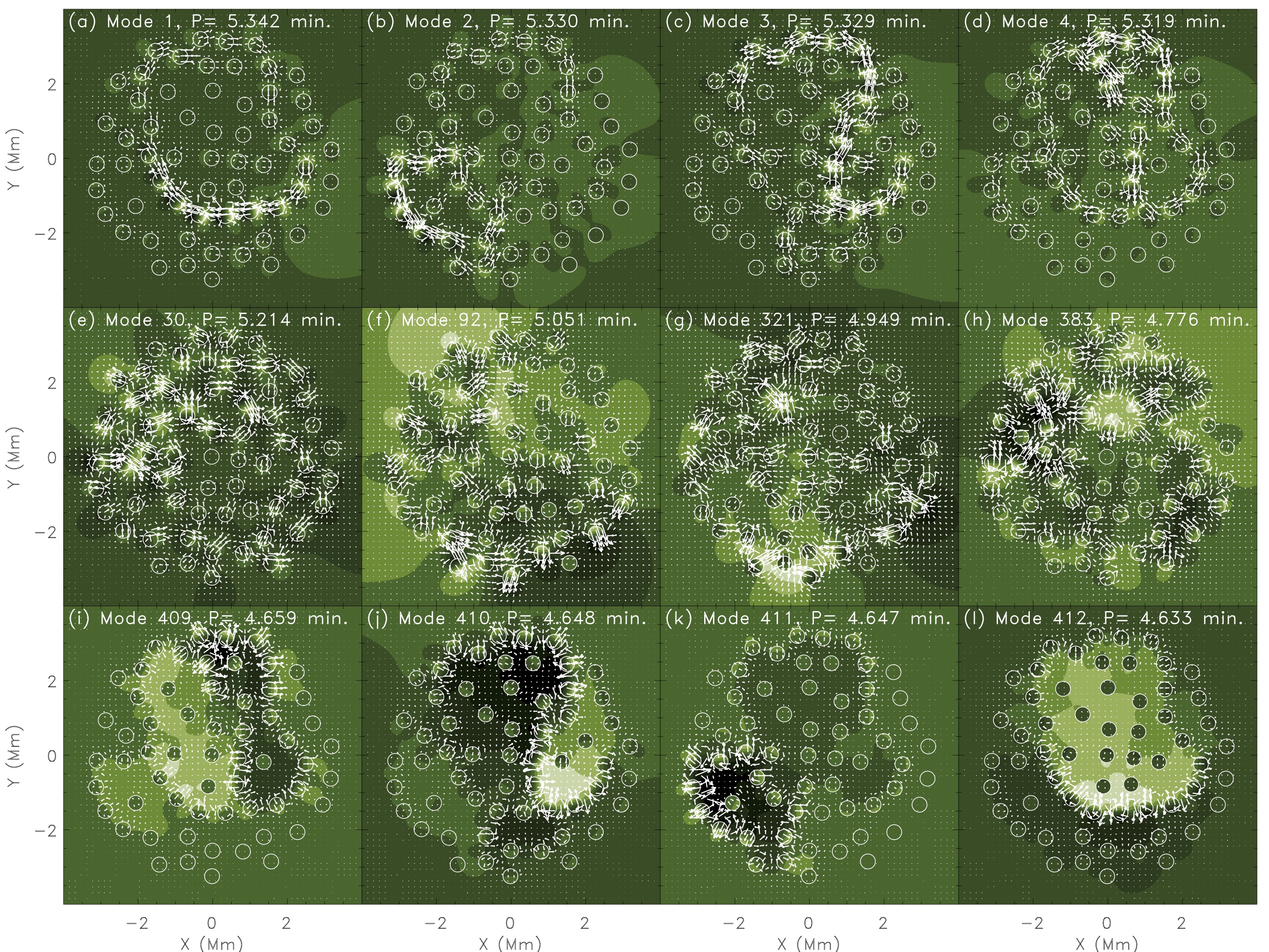} 
\caption{Total pressure perturbation (colour field) and velocity field (arrows) of the fast collective normal modes. From {\bf(a)} to {\bf(d)} corresponds to the first Low modes labelled from Mode 1 to 4. The central panels {\bf(e)}-{\bf (h)} are four examples of Mid modes. Panels {\bf (i)}-{\bf (l)} correspond to the last four High modes, Modes 409 to 412. \label{fig:multipanel-modes-60}}
\end{figure*}

In simple strand configurations with few closer elements, all the tubes participate in each collective mode. However, in more complex systems the situation is more involved. Figure \ref{fig:multipanel-modes-60} shows some of the normal modes of the system with 60 strands. The frequencies of the selected modes are marked with circles in the scatter period plot in Figure \ref{fig:periods60}. The top row (Figs. \ref{fig:multipanel-modes-60}(a)-\ref{fig:multipanel-modes-60}(d)) shows the first four Low modes and the bottom row (Figs. \ref{fig:multipanel-modes-60}(i)-\ref{fig:multipanel-modes-60}(l)) shows the last four High modes. From the figure, we can see that there are only subsets of tubes that participate in the collective oscillation. We have called this effect \emph{clustering} and the subsets of tubes participating in each mode are the \emph{clusters}.
There is a correspondence between the Low and High modes. From Figures \ref{fig:multipanel-modes-60}(a) and \ref{fig:multipanel-modes-60}(l) we see that both mode velocity fields are orthogonal. Similarly, the modes plotted in Figures \ref{fig:multipanel-modes-60}(b) and \ref{fig:multipanel-modes-60}(k) are also orthogonal. In general, for each Low mode, there is an orthogonal High mode. We can see this correspondence in all the modes plotted in Figure \ref{fig:multipanel-modes-60}. Thus a Low normal mode consists of a cluster of tubes oscillating in the direction that joins together the cluster elements. In contrast, the corresponding orthogonal High mode corresponds to motions perpendicular to that direction compressing and rarefying the coronal plasma. Additionally, in the mode with the highest period, $\plow$, the cluster of tubes that oscillates forms a \emph{chain} of tubes moving following one {another}. In the mode with the lowest period, $\phigh$, the same \emph{chain} of tubes oscillates but in the perpendicular direction. Figures \ref{fig:multipanel-modes-60}(e)-\ref{fig:multipanel-modes-60}(h) show some Mid mode examples. These modes are also perpendicular between them as the pairs of Low and High modes. For example, the mode shown in the panel \ref{fig:multipanel-modes-60}(e) is perpendicular to the case shown in \ref{fig:multipanel-modes-60}(h). Also both modes shown in \ref{fig:multipanel-modes-60}(f) and \ref{fig:multipanel-modes-60}(g). Mid modes have more spatial complexity and the structure is not clearly organized forming large clusters. In these modes it seems that a large number of strands participate in the collective motions. However, small subsets are formed resembling a non-organised motion. 





\section{Dependence of the collective oscillations with the size of the system}\label{sec:depencewithsystemsize}

In this section, we study the normal mode properties of systems with an increasing number of strands in order to understand the fundamental vibrations of the solar corona.
\begin{figure}[!ht]
\centering\includegraphics[width=0.49\textwidth]{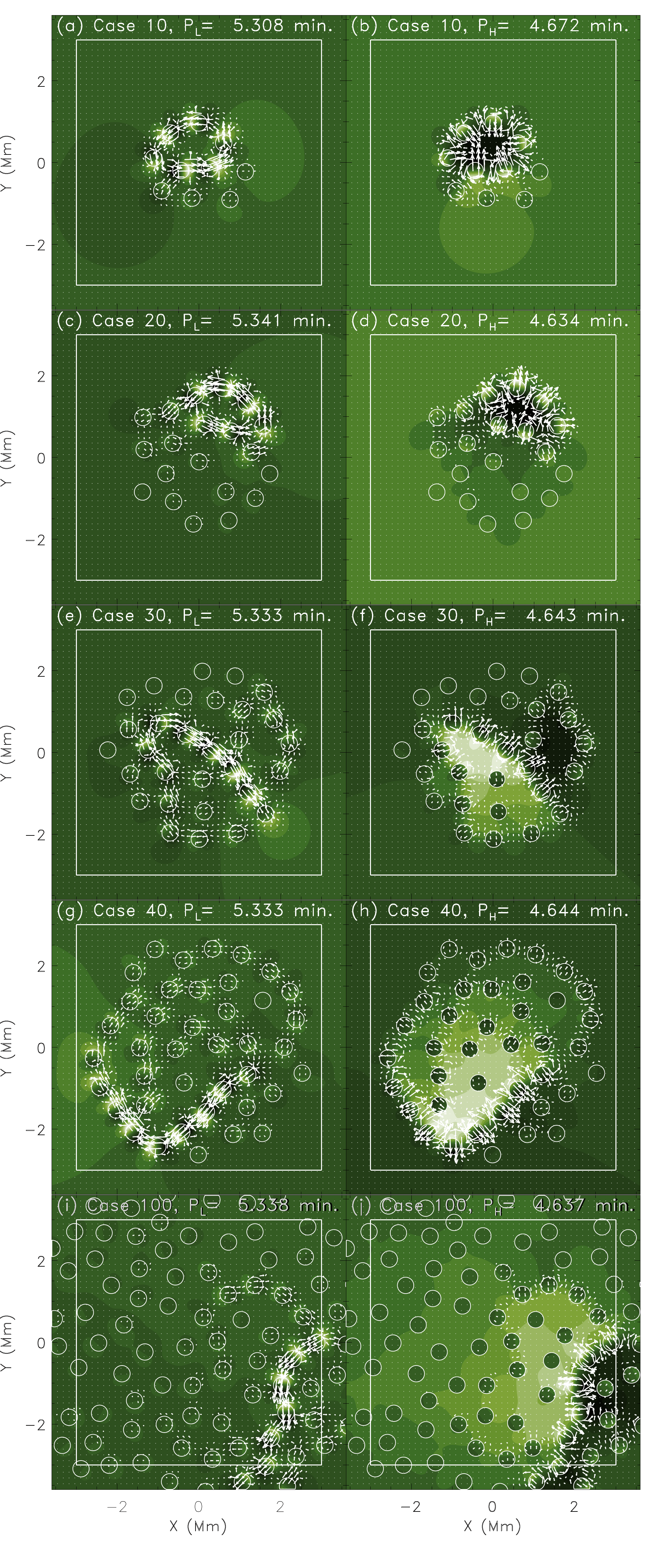}
\caption{Plot of the first and last normal modes for $\nstr=10, 20, 30, 30$ and 100 cases in each row. The left column panels are the first normal mode, associated with $\phigh$. The right column is the last normal mode, associated with $\plow$.     \label{fig:multipanel-modelowtopcomparison}}
\end{figure}
In previous Section \ref{sec:clustering} we have found that not all the strands participate in each normal mode and only a \emph{cluster} of strands oscillates. Figure \ref{fig:multipanel-modelowtopcomparison} shows the first and last normal modes with periods $\plow$ and $\phigh$ respectively for $\nstr=10, 20, 30, 50$ and 100 cases. In all the cases the oscillations are in the form of \emph{clusters} forming a \emph{chain} of strands. In both modes, the oscillation is in the same chain of tubes. However, both Low and High modes are mutually perpendicular as discussed in Section \ref{sec:clustering}. These chains can be closed (Figs. \ref{fig:multipanel-modelowtopcomparison}(a) and \ref{fig:multipanel-modelowtopcomparison}(c)) or open (Fig. \ref{fig:multipanel-modelowtopcomparison}(e), \ref{fig:multipanel-modelowtopcomparison}(g) and \ref{fig:multipanel-modelowtopcomparison}(i)). In all the situations, the normal modes associated to $\plow$ and $\phigh$ are chains of the nearest set of strands with a separation between them close to $d_\mathrm{min}=\mathcal{S} \, 2 \, \rstr= 0.68 \Mm$ (see examples in Fig. \ref{fig:multipanel-modelowtopcomparison}). {The rest of Low and High modes also form clusters but due to space limitations they are not shown in Figure \ref{fig:multipanel-modelowtopcomparison}.  In order to estimate the number of tubes forming a cluster in each normal mode we define the function 
\begin{equation}\label{eq:clustersizefunction}
    S_\mathrm{cluster}=\frac{A_\mathrm{s}(50\%)}{A_\mathrm{s}} \, \nstr \, ,
\end{equation}
where $A_\mathrm{s}(50\%)$ is the cross-section area inside the tubes with a velocity equal or larger than 50\% of the maximum velocity of the normal mode. $A_\mathrm{s}$ is the cross-sectional area of all the tubes. This function (\ref{eq:clustersizefunction}) is a good approximation of the number of tubes in a cluster. Figure \ref{fig:cluster-sizes} shows $S_\mathrm{cluster}$ as function of the mode periods for $\nstr=$ 10, 20, 30, 60, 100 cases. The shaded area covers the frequency range of the Mid modes for $\nstr=$ 60 and 100 cases (see Fig. \ref{fig:multipanel-modes-60}(e)-\ref{fig:multipanel-modes-60}(h)).
\begin{figure*}[!ht]
\centering\includegraphics[width=0.99\textwidth]{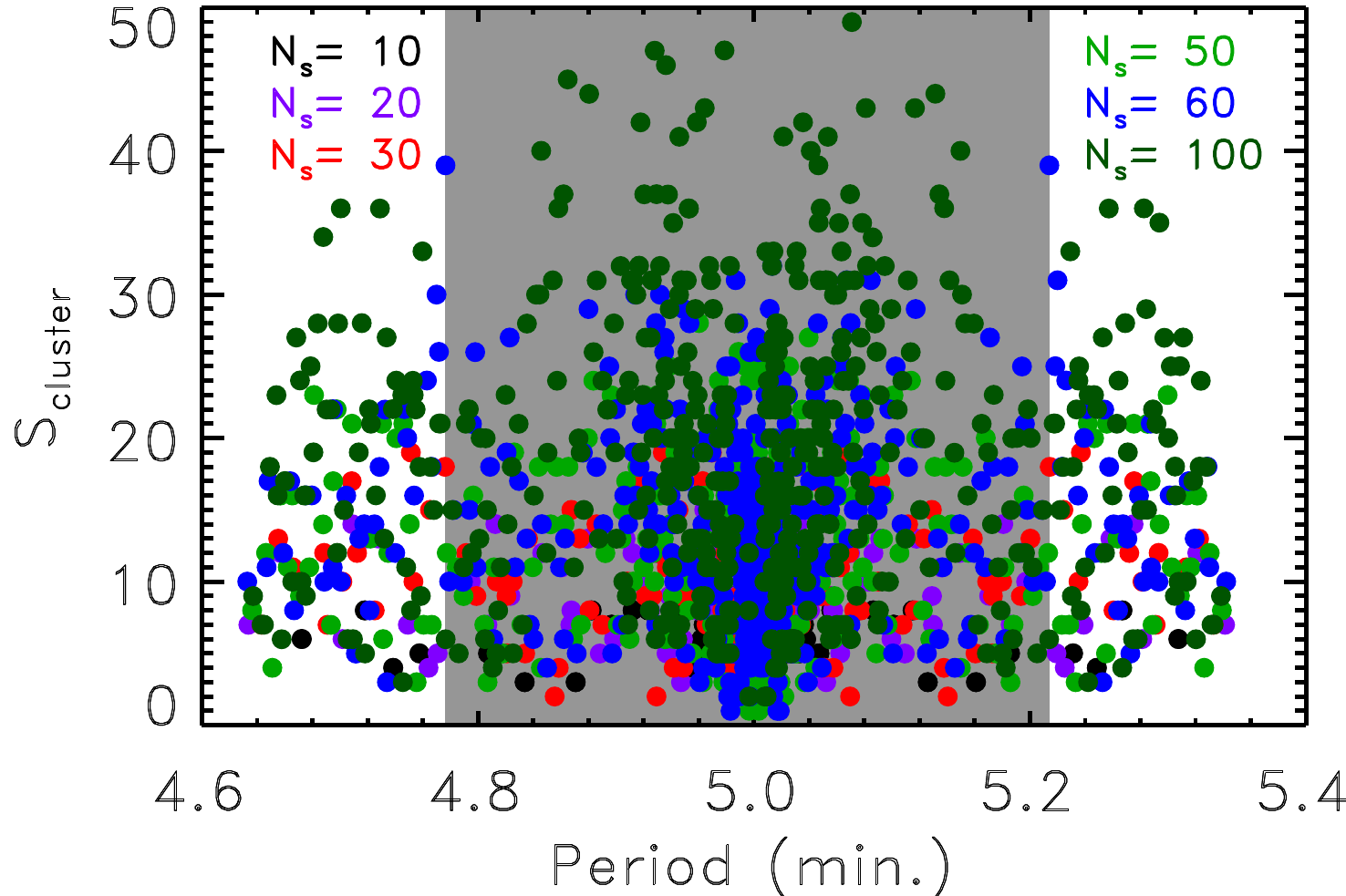}
\caption{Scatter plot of the approximate sizes of the cluster of strands of each normal mode given by the function $S_\mathrm{cluster}$ as a function of the mode periods for $\nstr=$ 10, 20, 30, 60, 100 cases in black, magenta, red, green, dark blue and dark green respectively. The shared are corresponds to the period range of the Mid modes for $\nstr=100$.\label{fig:cluster-sizes}}
\end{figure*}
The first and last modes, associated with $\phigh$ and $\plow$, form chains with less than 15 strands in all the cases consistent with Figure \ref{fig:multipanel-modelowtopcomparison}. This suggests a trend for $\nstr>100$, $S_\mathrm{cluster}$ may be also smaller than 15. We have computed the normal modes of a system of 200 strands confirming this tendency. However, due to numerical reasons, we cannot find the Mid modes and the results show a large gap of modes in the central periods close to $\pstr$. These chains are associated with the groups of closest tubes, i.e., $d\approx d_\mathrm{min}$, indicating that the size of the chains depends on the particular distribution of the tubes. In contrast, for the rest of Low and High modes the size of the clusters is much larger. From this figure, it is not clear if there is a general trend between the number of strands of the system and the size of the cluster. The magnitude $S_\mathrm{cluster}/\nstr$ seems to stabilise and it is more or less identical for 60 and 100 tubes. However, it is necessary to study much larger systems to find a general trend of $S_\mathrm{cluster}/\nstr$.}

In all the situations with $\nstr < 100$ the system is inside the ROI of $6\times6 \Mm^2$ plotted as a white square area in Figure \ref{fig:multipanel-modelowtopcomparison} and hence the clusters are inside the area. However, for $\nstr =100$ both clusters plotted in \ref{fig:multipanel-modelowtopcomparison}(i) and \ref{fig:multipanel-modelowtopcomparison}(j) the motion is partially outside the area. { From Figure \ref{fig:cluster-sizes} we see that many normal modes of this system have much larger clusters. Thus, many normal modes involve strands inside and outside the ROI. In a much larger system with $\nstr> 100$ there will be even larger clusters partially inside and outside the area. The modes with clusters partially inside indicate the coupling of the vibrations of the ROI strands with the surrounding corona. In this sense, the range of the coupling of the strands of ROI extends beyond a system of 100 strands. Thus, the system of 100 strands is not fully representative of the oscillations in the ROI.}



\begin{figure*}[!ht]
\centering\includegraphics[width=0.9\textwidth]{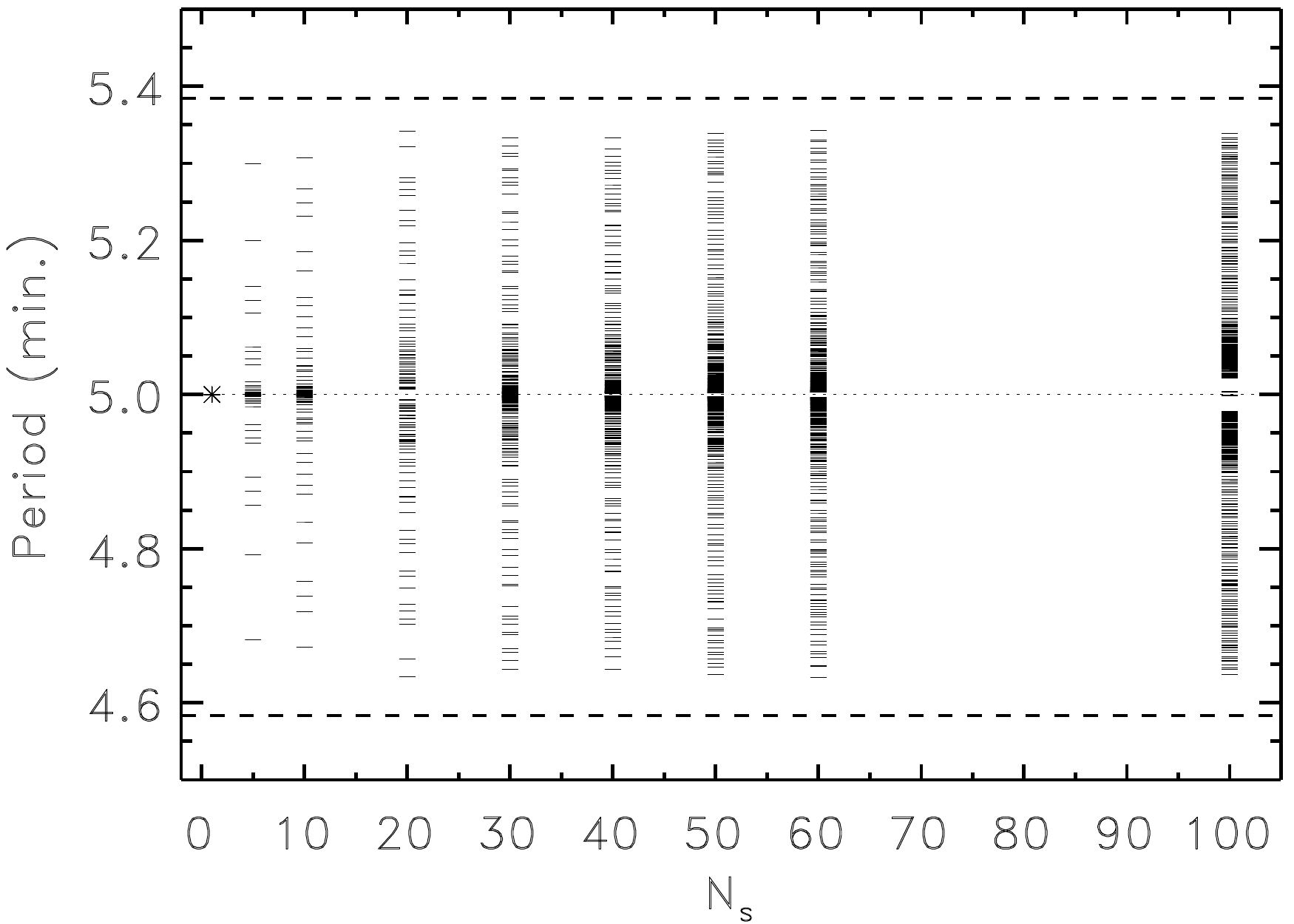}
\caption{Scatter plot of the normal mode periods of the systems considered a function of the number of strands, $\nstr= $ 1, 5, 10, 20, 30, 40, 50, 60 and 100. Each period is drawn as a small horizontal line except the case of $\nstr=1$ where an asterisk is used. The horizontal dotted line is the period of oscillation of an individual strand, $\pstr = 5 \min$. Both horizontals dashed lines are the analytic approximate expressions for $\phigh$ and $\plow$ from Eq. (\ref{eq:periods}). In all the situations the number of modes increases with the proximity to $\pstr$ and the plot lookslike a continuous dark band. In the case of 100 strands, there is a gap associated with a numerical issue in the T-method. Near the central frequency, the T-matrix becomes singular and the method fails near the singularity.}\label{fig:periods_identical_strands}
\end{figure*}

Figure \ref{fig:periods_identical_strands} shows the normal mode periods for all the considered systems with $\nstr=$1, 5, 10, 20, 30, 40, 50, 60 and 100 strands.
The period band, $\Delta P$, increases rapidly between 1 to 10 strands but for $\nstr\ge20$, $\plow$ and $\phigh$ reach an approximately constant value. With this plot, we obtain a relevant result: the period band or period splitting is independent of the size of the system {for a sufficiently large $\nstr$}. In this sense, we obtain a general result that can be applied to an unlimited set of strands. The reason for this effect is that the two modes that flank the period band, i.e. the modes with $\phigh$ and $\plow$, are chains of close tubes with a separation between them similar to $d_\mathrm{min}$. In this sense, the period band depends on the random distribution of the strands and not on the size of the system.

Additionally, the number of modes increases with $\nstr$. For $\nstr=5$ there are a few scattered modes in the period band whereas for $\nstr=100$ they almost form a continuum of modes. The reason is that the number of clusters increases with the number of strands. In fact, for an infinite number of strands, the periods will tend to form a continuum between $\phigh$ and $\plow$. However, these modes are associated with clusters of strands in different spatial locations. {Considering a finite volume, not all the normal modes have clusters inside the region and most of them are outside the finite volume}.
Therefore, the oscillations in a ROI will consist of a bunch of collective modes forming a very densely populated band of periods, but not a continuum. 

For the ROI considered here, shown in Figure \ref{fig:model}, the normal modes of the $\nstr = 100$ system are not a full description of the oscillations. Many collective normal modes are missing, with clusters involving tubes of the ROI and extending much beyond this region. However, due to the difficulty of numerically handling systems with a huge number of strands, we believe that the $\nstr = 100$ case as a good qualitative representation of the vibrations of the ROI in terms of the obtained period band and complexity of the strand motions.





\subsection{Analytic approximation of period splitting}\label{analyticapproximationperiodsplitting}
The normal mode with the highest period, $\plow$, is formed with a chain of strands following one another (Fig. \ref{fig:multipanel-modes-60}(i)). Additionally, in the normal mode with the lowest period, $\phigh$, the same chain of strands oscillates in the perpendicular direction (Fig. \ref{fig:multipanel-modes-60}(j)). The periods of both modes define the period splitting associated with the strands interaction, $\Delta \, P = \plow -\phigh$, i.e. without interaction $\Delta \, P = 0$.
\begin{figure}[!ht]
\centering\includegraphics[width=0.45\textwidth]{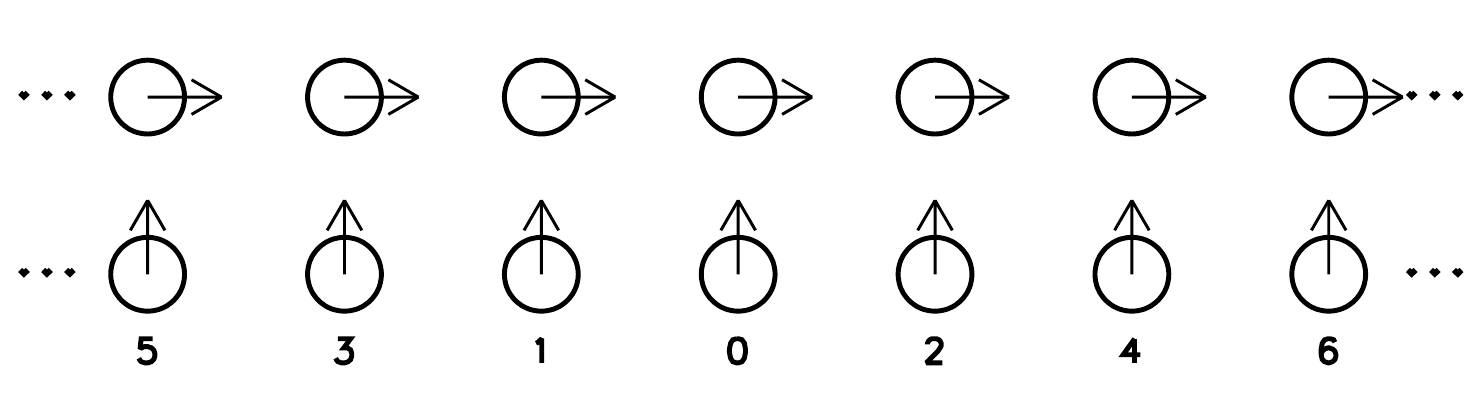}
\caption{Sketch of the normal modes associated with $\plow$ (top) and $\phigh$ (bottom) of an infinite array of aligned strands along the horizontal axis. The strands are equally spaced with a separation $d$ between the strands. Arrows indicate the direction of the motion. In the top panel, the strands move along the horizontal axis and in the bottom panel, the strands move perpendicular to this axis. 
The three dots at both ends means that the system continues to the infinity in both directions. The numbers indicate how the strands are labelled.}\label{fig:infinite-array-model}
\end{figure}
In order to find an approximate expression for $\plow$ and $\phigh$ we approximate the chain of strands as an infinite array of aligned tubes. In Figure \ref{fig:infinite-array-model} the two modes with the highest period (top) and lowest period (bottom) are shown. From \citet{luna2009} the total pressure perturbation, $ \psi$, can be expressed as
\begin{equation}\label{eq:netfield_ex+sc}
 \psi(\mathbf{r})=\sum_{\mathrm{m}=-\infty}^{\infty}
    \alpha_\mathrm{m}^\mathrm{j} \left[  J_\mathrm{m} (k_\mathrm{0}
    R_\mathrm{j}) + T_\mathrm{m m}^\mathrm{j}  H^{(\mathrm{1})}_\mathrm{m}
    (k_\mathrm{0} R_\mathrm{j}) \right] e^{i \mathrm{m} \varphi_\mathrm{j}} \, ,
\end{equation}
for the external medium to the strands where 
\begin{equation}
k_\mathrm{0}^{2} = \frac{\omega^{2}-k_{z}^{2}v_\mathrm{A0}^{2}}{v_\mathrm{A0}^{2}}\, ,
\end{equation}
{$v_\mathrm{A0}^{2}=B^{2}/\mu\rho_\mathrm{c}$, $J_\mathrm{m}$ and $H^{(\mathrm{1})}_\mathrm{m}$ are the Bessel and Hankel functions of the first kind and order $m$ and $T_\mathrm{m m}^\mathrm{j}$ are the T-matrix elements, that represent the scattering properties of the $j$-th tube \citep[see Eq. 17 of][for details]{luna2009}}.
This field has two contributions associated with the excitation field to the $j$-tube and the scattered field by the same tube. The excitation field is the scattered field by the other tubes {that drives the oscillations of the $j$-tube}. The coefficients $\alpha_\mathrm{m}^\mathrm{j}$ are the expansion coefficients of order $m$, that depend on the $k_z=\pi/L$ wave number and the angular frequency $\omega$, and $R_\mathrm{j}$ and $\varphi_\mathrm{j}$ are the local polar coordinates centered at $\mathbf{r}_\mathrm{j}$, defined through $R_\mathrm{j}=|\mathbf{r}-\mathbf{r}_\mathrm{j}|$ and $\cos \varphi_\mathrm{j} =\mathbf{e}_x \cdot (\mathbf{r}-\mathbf{r}_\mathrm{j})/|\mathbf{r}-\mathbf{r}_\mathrm{j}|$. As explained in \citet{luna2009} it is possible to find an equation for the $\alpha_\mathrm{m}^\mathrm{j}$ coefficients and mode frequencies $\omega$. This equation is
\begin{equation}\label{eq:problem_solution}
\alpha_\mathrm{m}^\mathrm{j}+\sum_ {\mathrm{q} \neq \mathrm{j}}^{\mathrm{N}}\sum_{\mathrm{n}=-\infty}^{\infty} \alpha_\mathrm{n}^\mathrm{q} T_\mathrm{n n}^\mathrm{q}
H_\mathrm{n-m}^{(\mathrm{1})}(k_\mathrm{0}|\mathbf{r}_\mathrm{j}-\mathbf{r}_\mathrm{q}|) e^{i (\mathrm{n}-\mathrm{m}) \varphi_\mathrm{j q}} =0 ,
\end{equation}
where $\varphi_\mathrm{j q}$ is the angle formed by the center of the $i$-th loop with respect to the center of the $j$-th flux tube. Equation (\ref{eq:problem_solution}) is formally an infinite system of equations for an infinite number of unknowns ($\alpha_\mathrm{m}^\mathrm{j}$). The condition that there is a nontrivial solution, i.e., the determinant formed by the coefficients set equal to zero, provides us with the dispersion relation \citep[see details in][]{luna2009,luna2010}. For practical reasons the equation is approximated by truncating the system to a finite number of equations and unknowns by setting $\alpha^\mathrm{j}_{\mathrm{m}+1}=0$ for azimuthal numbers greater than a truncation number ($m>m_\mathrm{t}$). To ensure the convergence of solutions, they must be independent of the truncation number $m_\mathrm{t}$. However, in this section in order to find an approximate dispersion relation, we assume that both modes of Figure \ref{fig:infinite-array-model} are exclusively kink-like modes with the interaction of the strands through $m=\pm 1$.



Both Equations (\ref{eq:problem_solution}) for the central strand of Figure \ref{fig:model} labeled 0 for $m=\pm 1$ are
\begin{eqnarray}\nonumber
\frac{\alpha_{1}^{0}}{T} &+& \sum_{p=1}^{N}\left[  H_{0}(k_\mathrm{0} \, p\, d)\left\{ \alpha_{1}^{2 p-1} +\alpha_{1}^{2 p} \right\}+ \right. \\ \label{eq:system1}
&& ~~~~ \left.+  H_{2}(k_\mathrm{0} \, p\, d)\left\{ \alpha_{-1}^{2 p-1} +\alpha_{-1}^{2 p} \right\} \right]=0 \\ \nonumber
\frac{\alpha_{-1}^{0}}{T} &+& \sum_{p=1}^{N}\left[  H_{2}(k_\mathrm{0} \, p\, d)\left\{ \alpha_{1}^{2 p-1} +\alpha_{1}^{2 p} \right\}+ \right. \\ \label{eq:system2}
&& ~~~~ \left.+  H_{0}(k_\mathrm{0} \, p\, d)\left\{ \alpha_{-1}^{2 p-1} +\alpha_{-1}^{2 p} \right\} \right]=0
\end{eqnarray}
where $T \equiv T_{-1 -1}=T_{1 1} $, $H_\mathrm{2}(x) \equiv H_\mathrm{2}^{(\mathrm{1})}(x)=H_\mathrm{-2}^{(\mathrm{1})}(x) $ and  $H_\mathrm{0}(x) \equiv H_\mathrm{0}^{(\mathrm{1})}(x)$. Additionally $|\mathbf{r}_\mathrm{j}-\mathbf{r}_\mathrm{s}|=p \, d$ being $p=1, 2, 3, \dots$ and $e^{i (\mathrm{n}-\mathrm{m}) \varphi_\mathrm{j i}}=1$ in all the combinations of $m$ and $n$. In these equations we see that the strand 0 interacts with strands 1 and 2 that are at a distance $d$, with strands 3 and 4 at a distance $2d$ and in general with a couple of strands at a distance $p \, d$. In principle $N = \infty$ but the interaction between strands is negligible at some distance larger than $N d$. {The reason is that the coupling between two strands is given by the functions $H_{0}(x)$ and $H_{2}(x)$ that are proportional to $1/x^2$ for an small $x$}. In a general situation it is necessary to consider the equivalent equations to (\ref{eq:system1}) and (\ref{eq:system2})  for all the strands solving the infinite equations system. However, we are interested in a situation where all the strands oscillate in the same way (see Fig. \ref{fig:infinite-array-model}) where
\begin{eqnarray}
&\alpha& \equiv \alpha_{1}^{0} ~~ =\alpha_{1}^{1}~~=\alpha_{1}^{2}~~=\dots=\alpha_{1}^{N} \, ,\\
&\beta& \equiv \alpha_{-1}^{0}=\alpha_{-1}^{1}=\alpha_{-1}^{2}=\dots=\alpha_{-1}^{N} \, ,
\end{eqnarray}
and the system reduces to two unknowns, i.e. $\alpha$ and $\beta$ and two equations,
\begin{eqnarray}\nonumber
\left[ \frac{1}{T} + \sum_{p=1}^{N}2 \, H_{0}(k_\mathrm{0} \, p\, d) \right] \alpha + \sum_{p=1}^{N}2 \, H_{2}(k_\mathrm{0} \, p\, d) \beta &=&0 \\ \nonumber
 \sum_{p=1}^{N} 2 H_{2}(k_\mathrm{0} \, p\, d) \alpha+ \left[ \frac{1}{T} +\sum_{p=1}^{N}2  H_{0}(k_\mathrm{0} \, p\, d)\right] \beta &=&0
\end{eqnarray}
for both $\alpha$ and $\beta$ coefficients. We obtain the dispersion relation from the condition that there is a nontrivial solution of the system, namely
\begin{equation}\label{eq:determinant}
\left[ \frac{1}{T} + \sum_{p=1}^{N}2 \, H_{0}(k_\mathrm{0} \, p\, d) \right]^2 - \left[ \sum_{p=1}^{N} 2 H_{2}(k_\mathrm{0} \, p\, d) \right]^2 =0 \, ,
\end{equation}
or
\begin{equation}\label{eq:dr}
 \frac{1}{T} + 2 \sum_{p=1}^{N} \left[ H_{0}(k_\mathrm{0} \, p\, d) \pm   H_{2}(k_\mathrm{0} \, p\, d) \right]  =0 \, ,
\end{equation}
where the $\pm$ comes from the square root of previous Equation (\ref{eq:determinant}). The interaction of the strands is constrained by their neighbours. In this sense the term $k_\mathrm{0} \, p\, d$ is small and $k_\mathrm{0} \, \rstr \ll 1$. Under this approximation the combination of Hankel functions in Equation (\ref{eq:dr}) is
\begin{equation}\label{eq:approx-hankel-expressions}
H_{0}(k_\mathrm{0} \, p\, d) \pm   H_{2}(k_\mathrm{0} \, p\, d) \approx \mp \frac{4 i}{\pi k_{0}^2 \, p^{2} d^{2}} \, ,
\end{equation}
and the T-matrix element can be approximated as
\begin{equation}
T=T_{1 1} \approx -\frac{\rho_{i}(\omega^{2}-k_\mathrm{z}^{2}v_\mathrm{A i}^{2})-\rho_{e}(\omega^{2}-k_\mathrm{z}^{2}v_\mathrm{A e}^{2})}{\rho_{i}(\omega^{2}-k_\mathrm{z}^{2}v_\mathrm{A i}^{2})+\rho_{e}(\omega^{2}-k_\mathrm{z}^{2}v_\mathrm{A e}^{2})}\frac{i \pi}{4} k_\mathrm{0}^{2} \, \rstr^{2} \, 
\end{equation}
\citep[see,][for details]{soler2015}. Inserting these approximated expressions into Equation (\ref{eq:dr}) we obtain the period of both modes as
\begin{equation}\label{eq:periods}
P=\pstr \sqrt{1 \pm \frac{\rho_\mathrm{s}-\rho_\mathrm{c}}{\rho_\mathrm{s}+\rho_\mathrm{c}} \, \mathcal{F} \,\left(\frac{\rstr}{d}\right)^{2}} \, ,
\end{equation}
where the $+$ and $-$ corresponds to $\plow$ and $\phigh$ respectively and
\begin{equation}\label{eq:factor}
\mathcal{F} = 2 \sum_{p=1}^{N} \frac{1}{p^{2}} \le \frac{\pi^{2}}{3} \, .
\end{equation}
In this summation, $N$ should be finite in order to keep $k_\mathrm{0} \, N\, d$ relatively small and the approximation (\ref{eq:approx-hankel-expressions}) valid. In addition, the interaction is restricted to few neighbouring tubes. However, considering $N\to \infty$, Equation (\ref{eq:factor}) converges to $\pi^{2}/3$. Thus, the parameter is always $\mathcal{F}<\pi^{2}/3$. We can consider that $\mathcal{F}=\pi^{2}/3$ as a good approximation {for large systems} and that the period splitting $\Delta P$ is always constrained to lower values than those given by Equation (\ref{eq:periods}). With the values of the parameters described in Sections \ref{sec:model} and \ref{sec:clustering} and assuming that $d=d_\mathrm{min}$, we obtain $\phigh=4.58$ and $\plow= 5.38 \min$. In Figure \ref{fig:periods_identical_strands}, we have plotted these values as two horizontal dashed lines. We see that for a sufficiently large number of strands the maximum and minimum periods are well approximated by these analytic expressions. Expression (\ref{eq:periods}) is very similar to Equation (30) by \citet{soler2015} but with a correction factor, $\mathcal{F}$, associated to the multiple interaction.


Assuming that the period splitting is approximately symmetric, $\pstr\approx1/2 (\plow+\phigh)$ and with Equation (\ref{eq:periods}) we obtain
\begin{equation}
\Delta P = \plow-\phigh=\pstr\frac{\rho_\mathrm{s}-\rho_\mathrm{c}}{\rho_\mathrm{s}+\rho_\mathrm{c}} \, \mathcal{F} \,\left(\frac{\rstr}{d}\right)^{2} \, .
\end{equation}
With this expression we can see that the frequency splitting depends essentially on 
$d$, which is the minimum distance in the strand distribution, $d_\mathrm{min}$. A shorter $d$ involves a larger period splitting. It is important to note that $d$ is not directly related to the filling factor, $f$. As explained in Section \ref{sec:model} in a random distribution of strands, the separation between strands ranges from a minimum to a maximum value. Then the group of strands with the minimum separation between them, $d_\mathrm{min}$, determines the maximum period splitting, $\Delta P$.





\begin{figure*}[!ht]
\centering\includegraphics[width=0.99\textwidth]{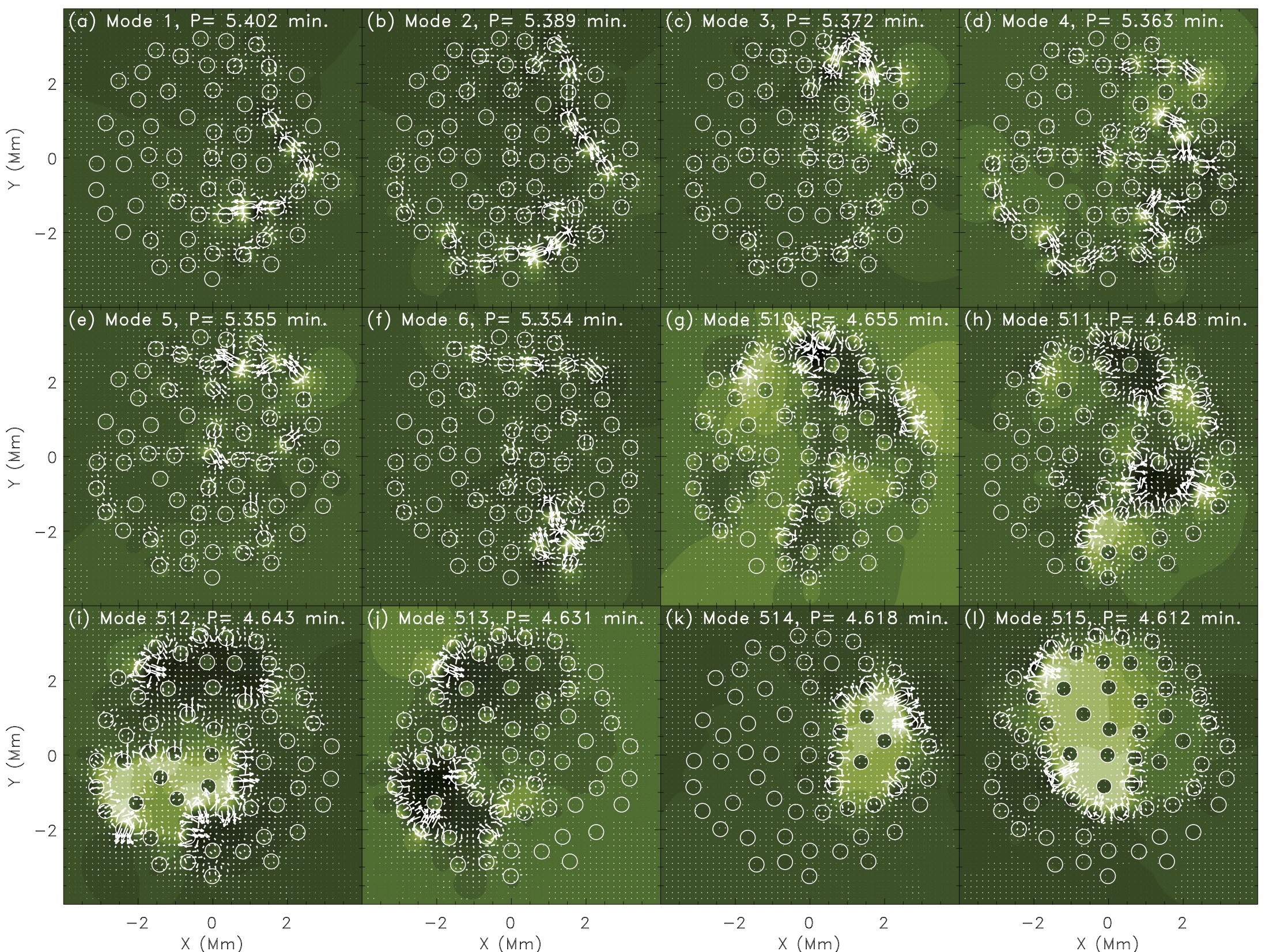}
\caption{Similar to Fig. \ref{fig:multipanel-modes-60} showing some normal modes of the identical strand distribution but with a random density distribution from $3.220 \rho_\mathrm{c}$ to $3.815 \rho_\mathrm{c}$. The first six Low modes are plotted in panels (a) to (f) whereas the last six High modes are in panels (g) to (l).\label{fig:multipanel-modes-60-nonidentical}}
\end{figure*}
\section{Non-identical tubes}\label{sec:non-identicalstrands}
The aim of this work is to understand the vibrations of a large region of the AR corona. The strands within this region have different tube lengths because of the 3-dimensional topology of the field. In \citet{luna2009} we found that differences in the strand Alfv\'en frequencies, i.e. $\kzva$, reduce or avoid the coupling between the strand oscillations \citep[see also,][]{soler2009}. Thus a difference in strand lengths may result in a decoupling of the AR strands. We can estimate the strand length differences by assuming semicircular shapes of the field lines in the region. The central tube has a length $\pi L$ and the tubes of the periphery of the region $ \pi L (1 \pm \epsilon)$. Considering the ROI shown in Figure \ref{fig:multipanel-modes-60} and with $L=100\Mm$, $\epsilon=0.04$ (i.e. 4\% of $L$). Properly accounting for the differences in tube lengths involves a fully 3-dimensional study, but our approach is restricted to 2.5-dimensions assuming that all the field lines have identical lengths. However, we can mimic the strand length differences effect by considering density differences. The relevant factor in strand coupling is the Alfv\'en frequency, $\kzva$. Thus a variation in length, $ \pi L (1 \pm \epsilon)$ is equivalent to a density variation $\rho \, (1 \pm \epsilon)^{2}$ in order to keep the variation in the Alfv\'en frequency the same,
\begin{equation}
\frac{\pi}{\pi L (1 \pm \epsilon)} \frac{B}{\sqrt{\mu_{0} \, \rho}} = \frac{\pi}{\pi L } \frac{B}{\sqrt{\mu_{0} \, \rho (1 \pm \epsilon)^{2}}} \, .
\end{equation}
In addition, for a small $\epsilon$,  $(1 \pm \epsilon)^{2} \approx 1 \pm 2 \, \epsilon$. Thus in our situation, a length variation of 4\% is equivalent to a density variation of 8\%. This density variation has probably a well defined spatial distribution associated with the continuous variation of the field lines lengths. However, the distribution of tube lengths in the $xy$-plane will depend on the AR field topology. We do not know this field length distribution a priori and we assume a random density distribution spanning an 8\% variation around a fixed density.



Figure \ref{fig:multipanel-modes-60-nonidentical} shows the first and the last 6 modes of the case of 60 strands of Figure \ref{fig:model} but with a strand density distribution ranging from $3.220 \rho_\mathrm{c}$ to $3.815 \rho_\mathrm{c}$. We see that the modes are different when compared to those of the identical tubes shown in Figure \ref{fig:multipanel-modes-60}. For example, in the fundamental mode (Mode 1), the cluster of oscillating strands is different in both situations. The density variations in the strands affect the couplings between them and in general new clusters appear. {The correspondence we have found in Section \ref{sec:clustering} between Low and High modes is lost as we see in Figure \ref{fig:multipanel-modes-60-nonidentical}(a) and \ref{fig:multipanel-modes-60-nonidentical}(l)}. It is remarkable that in a situation with strand differences the coupling remains. The periods $\plow$ and $\phigh$ are similar to those of the identical tubes case. In this sense, the results of the analysis of Section \ref{sec:depencewithsystemsize} can be also applied here and the approximate expression given by Equation (\ref{eq:periods}) remains valid for non-identical strands with small tube length differences.



\section{Forward modelling}\label{sec:syntheticimages}
In this section, we generate synthetic images to compare with the observational evidence in a technique called forward-modelling. We consider the fundamental oscillations of the ROI and generate synthetic images with the FoMo code \citep[\url{https://wiki.esat.kuleuven.be/FoMo};][]{antolin2013,van-doorsselaere2016}. We focus on the fingerprints of these vibrations in the synthetic Dopplergrams and the non-thermal line broadening. The vibrations of the strands system is a linear combination of the normal modes as follows
\begin{equation}\label{eq:velocity-reconstruction}
\vec{v}(x,y,z,t) = \sum_{n=1}^{N_\mathrm{modes}}  A_{n} \, \hat{\vec{v}}_{n}(x,y,z) \, \cos(\omega_{n} \, t +\phi_{n} ) \,
\end{equation}
where $\hat{\vec{v}}_{n}$ is the $n$-th normal mode velocity normalised to unity. $\omega_{n}$ is the angular frequency of each normal mode. $A_{n}$ and $\phi_{n}$ are the amplitude and initial phase respectively of the velocity perturbation of the $n$-th mode.
\begin{figure}[!ht]
\centering\includegraphics[width=0.45\textwidth]{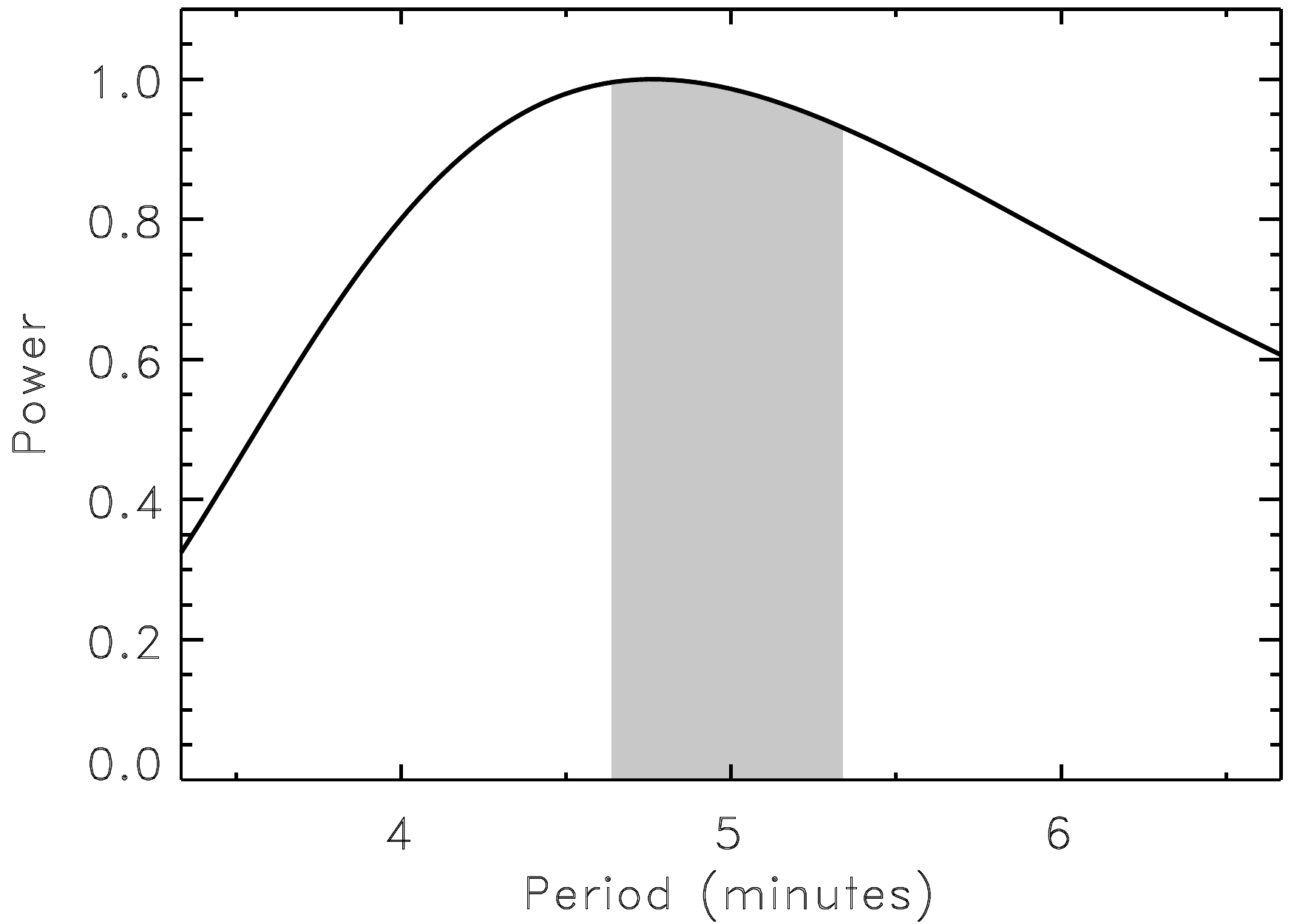}
\caption{Power spectra from \citet{tomczyk2007} normalized to its maximum (see text for details). The shaded area corresponds to the period range of the normal modes of the case of 100 strands shown in Fig. \ref{fig:periods_identical_strands}. \label{fig:power-spectra}}
\end{figure}
In general, the amplitude and phase of each normal mode will depend on how the system is perturbed. We assume that the oscillatory spectrum is similar to the one observed by \citet{tomczyk2007} that consists of a Gaussian function of width $\sigma/2 \pi=1$ mHz and centred at $\omega_{0}/2 \pi=3.5$ mHz shown in Figure \ref{fig:power-spectra} as function of the period. The shaded area corresponds to the period range of the normal modes for a $\nstr=100$  system. The period spectrum of the normal modes is centred at 3.3 mHz (5 min) that is slightly shifted from the observed 3.5 mHz and covers a small portion of the measured spectral width. 
An additional consideration is that most of the collective mode periods are concentrated around 5 min as we see in Figure \ref{fig:periods_identical_strands}. These are mostly Mid modes with a complex spatial structure.
This introduces a bias in the distribution of the power given by Equation (\ref{eq:velocity-reconstruction}) to the 5 min period. In order to compensate this effect we compute the frequency distribution of the normal modes, $N(\omega_{n})$, and then define the amplitude of each normal mode as
\begin{equation}
A_{n}=v_{0} \, \frac{\sqrt{\mathrm{e^{-\frac{(\omega-\omega_{0})^{2}}{2 \sigma^{2}}}}}}{N(\omega_{n})} \; ,
\end{equation}
where $v_{0}$ is a scaling parameter.  The velocities inside the tubes change with time and $v_{0}$ is set to have a maximum velocity inside the tubes of $10\kms$ during the temporal evolution. Thus, $v_{0}$ will depend on the system of strands considered and the details of the linear combination. For the case of 100 strands considered in this section $v_{0}= 1.4 \kms$.
%
We have defined the initial phase $\phi_{n}$ as a random number between 0 and $2 \pi$. The resulting complexity of the oscillations of the strands is not very sensitive to the selection of $\phi_{n}$. The reason is that the periods are very different producing important phase differences during the temporal evolution even for identical $\phi_{n}$.
\begin{figure}[!ht]
\begin{center}
\includegraphics[width=0.5\textwidth]{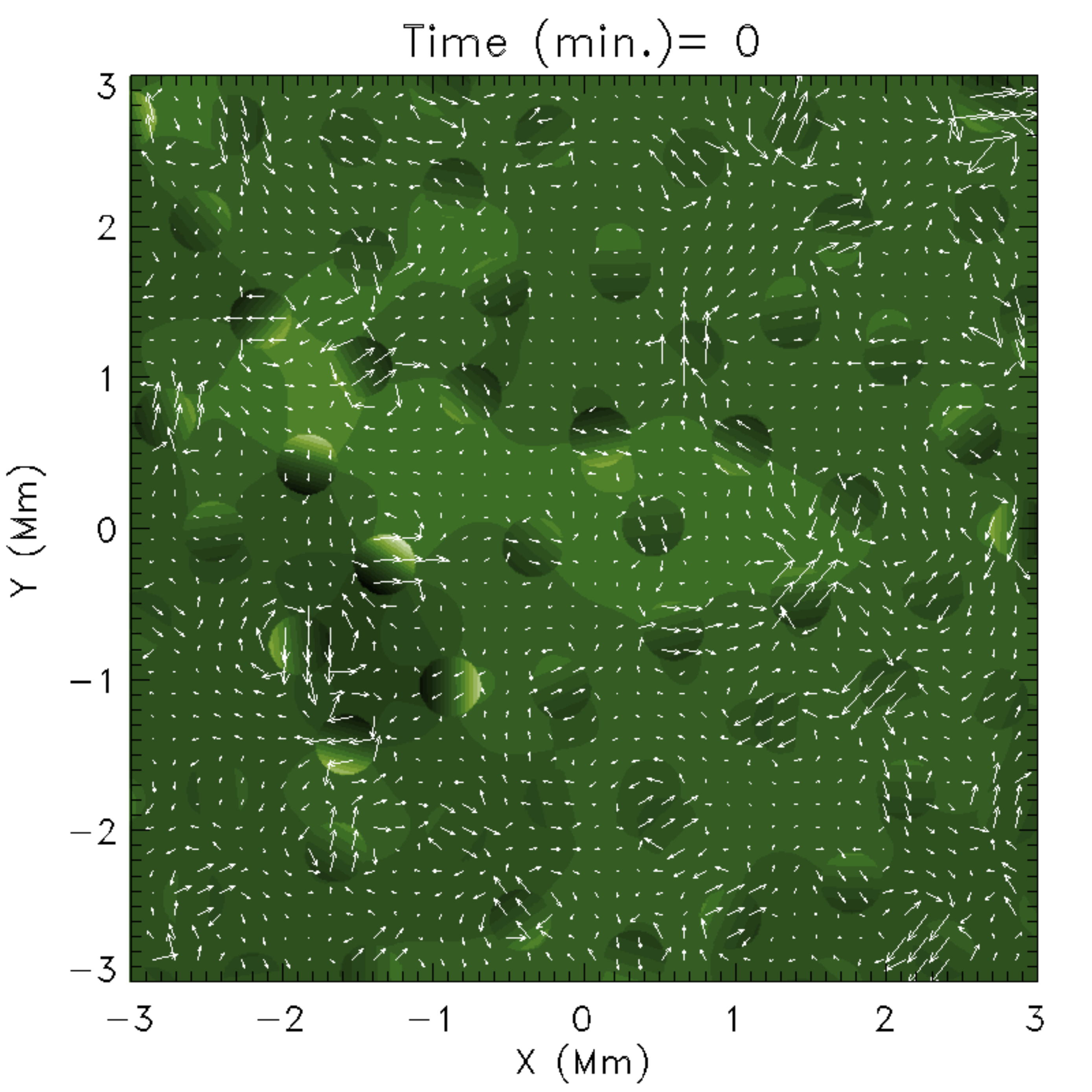}
\end{center}
\caption{First frame of the temporal evolution of the combination of normal modes of Eq. (\ref{eq:velocity-reconstruction}). Vectors are the velocity field and the green colour is the density perturbation. An animation of this figure is available in the online journal. \label{fig:movie}}
\end{figure}
Figure \ref{fig:movie} shows the temporal evolution of a linear combination of collective normal modes by Equation \ref{eq:velocity-reconstruction}. We clearly see the complex motions of the strands in the accompanying animation. In each strand, the kink oscillation direction changes with time. This change in the direction of the kink oscillation leads to clockwise or anticlockwise motions changing with time in a complex fashion. Additionally, the amplitude of the motion also changes producing amplification or damping of the motions. These effects were already reported by \citet{luna2010} with time-dependent numerical simulations. \cite{stangalini2017} have recently found a similar motion in chromospheric elements that can be understood as strand footpoints. In this sense, the helical motions in chromospheric tubes can be the result of the collective oscillations of the coronal strands. 

In order to generate the synthetic images, we need all the thermodynamic magnitudes: density, gas pressure and temperature. In our analytic calculations of the normal modes, we assume a zero-$\beta$ plasma. In this approximation, we obtain the velocity, magnetic field and density perturbations. The rest of the magnitudes are obtained with the linearised ideal MHD equations. We additionally assume that the equilibrium atmosphere is isothermal with $T_{0}=10^{6} $ K and the equilibrium gas pressure is computed with the ideal gas law and the equilibrium density defined in Section \ref{sec:model}. The gas pressure is not balanced at the strand surface between the internal and external medium. The reason is that we are in the zero-$\beta$ approximation and hence the gas pressure gradients are negligible in the equilibrium configuration. An additional assumption here is that we neglect the Lagrangian displacement of the plasma. Thus each LOS intersects always the same strands. However, considering the Lagrangian displacements, the strands intersected by each LOS change with time, introducing additional complexity in the synthetic Dopplergrams. Although a more exact treatment for the forward modelling would need this to be taken into account, in this work we are only interested in the qualitative picture that is produced in Dopplergrams by fundamental vibrations.

\begin{figure*}[!ht]
\begin{center}
\mbox{(a) \hspace{0.5\textwidth}(b)\hspace{0.5\textwidth}}\vspace{-0.5cm}
\centering\mbox{\includegraphics[width=0.5\textwidth]{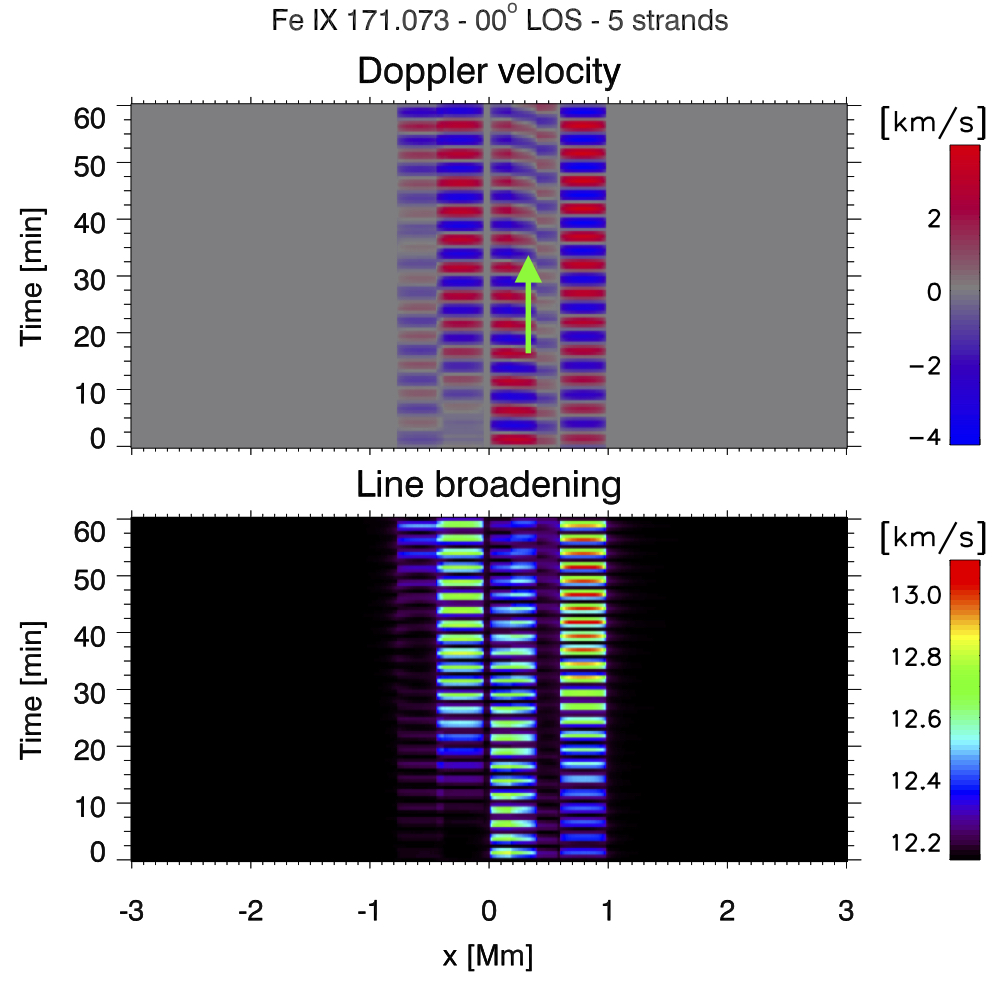}\includegraphics[width=0.5\textwidth]{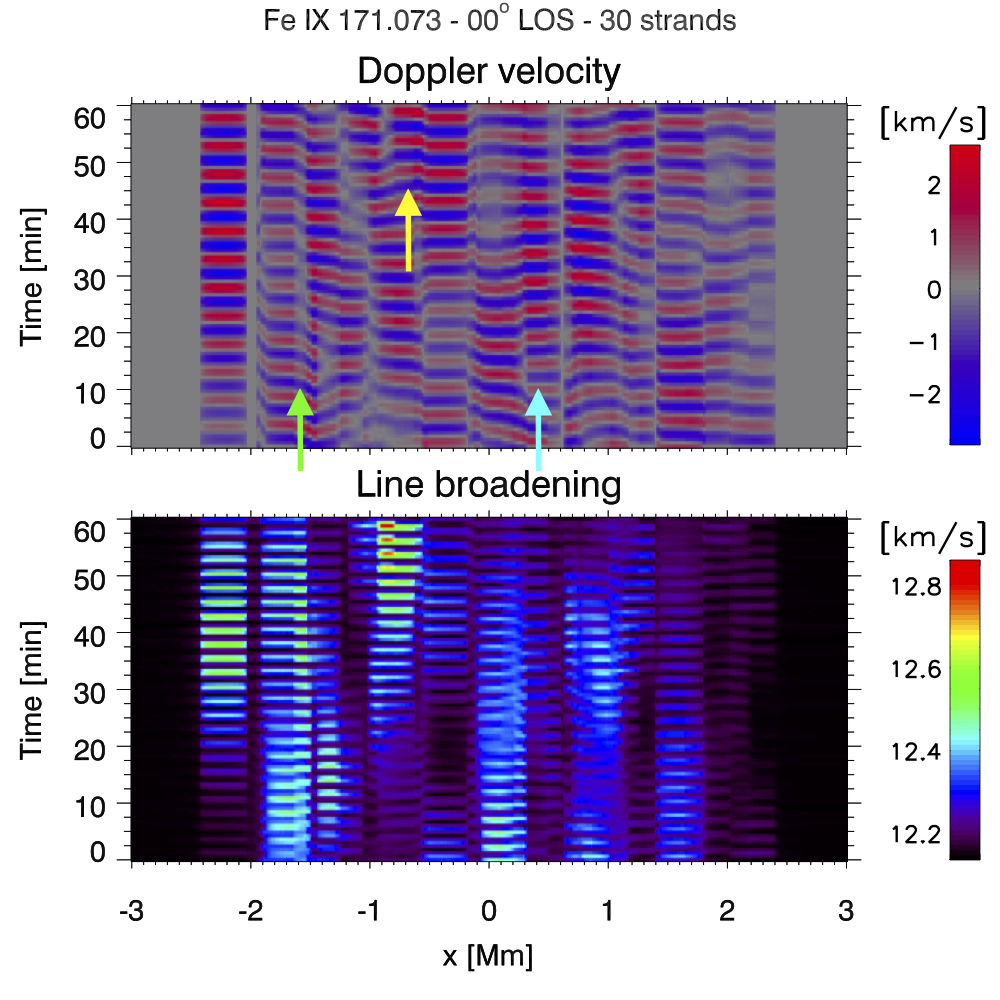}}
\mbox{(c) \hspace{0.5\textwidth}(d)\hspace{0.5\textwidth}}\vspace{-0.5cm}
\mbox{\centering\includegraphics[width=0.5\textwidth]{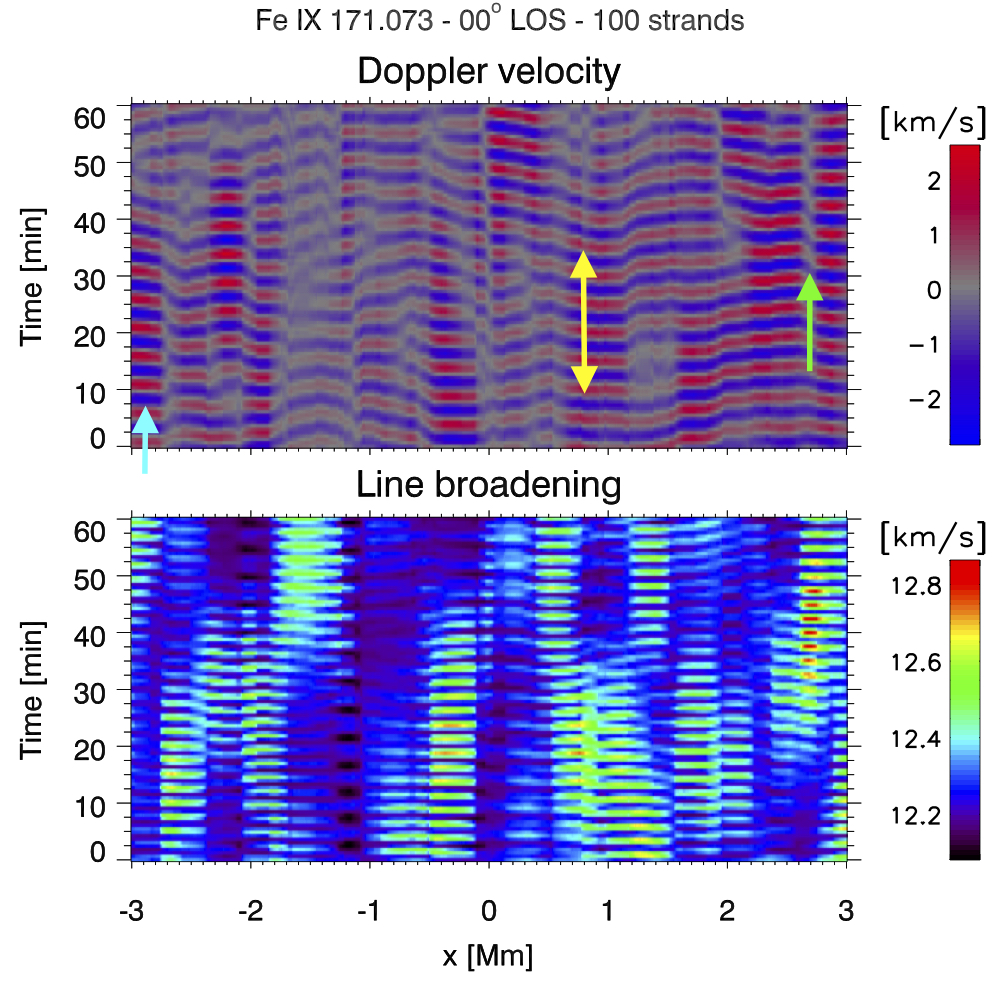}\centering\includegraphics[width=0.5\textwidth]{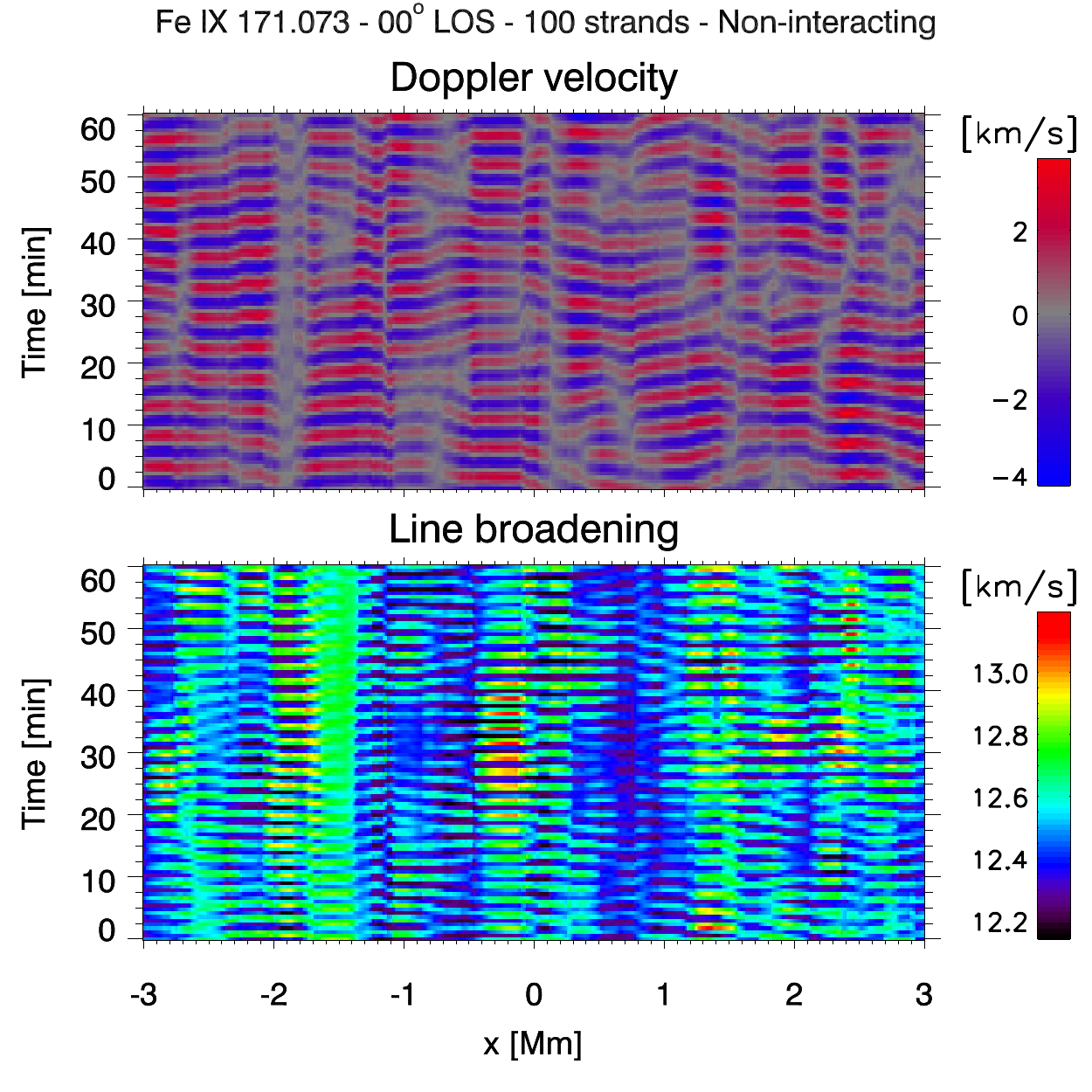}}
\end{center}
\caption{Synthetic images for $\nstr=100$ strands from the FoMo code for (a) 5 strands, (b) 30 strands and (c) 100 strands cases. In (d) a 100 non-interacting strands case is plotted (see text). Each panel shows time-distances of the Doppler and line broadening velocities. The arrows indicate some examples discussed in the text of regions with a constructive (light blue), destructive (green) or amplified of both Doppler and line broadening signals (yellow).
\label{fig:synthetic}}
\end{figure*}
For the forward modelling we choose the coronal spectral line of Fe IX 171.073\AA, forming at a temperature of 700,000 K approximately. We have also considered hotter coronal lines such as Fe XII 193.509\AA\ but the results do not vary much since the perturbations on the temperature produced by the fundamental vibrations are very small.  Therefore it is sufficient to only consider the results with the Fe IX line.

We consider an artificial slit placed perpendicularly to the strands direction and located at the loop apex. The artificial slit is placed along the $x$-axis in Figure \ref{fig:model} and covers the ROI region. The LOS is along the $y$-direction in the figure. Due to the line-tying of the magnetic field in the photosphere, the velocities of the strands are maximum at the apex of the tubes. Due to the finite size of our numerical box, any given LOS will cross more strands and less background corona for cases with larger number of strands. Since the inter-strand space contributes differently to the emergent intensity than the space within strands we have added dummy pixels at rest for each LOS ray to ensure that each ray has the same integrated length of background corona. 


Figure \ref{fig:synthetic} shows the time-distance diagrams on the artificial slit of the Doppler and line broadening signals for $\nstr=5, 30$ and 100. The 5 strand case is the most simplistic situation (Fig. \ref{fig:synthetic}(a)). In this case, there are one or two strands for each LOS. We can identify the two left and the rightmost individual tubes. In these positions, the Dopplergrams show the intrinsic motions of the individual strands projected along the  LOS. In the three tubes, we see a kink-like motion with a modulation of the amplitude. This modulation is associated with the interaction of the strands and it is a characteristic of the collective oscillation as we have discussed before (Fig. \ref{fig:movie}). This consists of a rotation of the direction of the oscillations in combination with damping or amplification of the motion. The maximum Doppler velocity is around $4\kms$ but the actual strand velocities have a maximum of $10\kms$. In these three cases the Doppler signal comes from individual tubes, which indicates that there is an intrinsic reduction of the observed velocity associated with the measurement \citep[see,][]{antolin2013}. This intrinsic reduction is less than a factor 3 with respect to the actual velocity. The line broadening is in phase with the Doppler signal. An increase of the Doppler velocity involves an increase of the line broadening. This coincidence indicates that the Doppler signal is associated with the actual motions of the strands. The line broadening has an almost uniform value between 12 and $14\kms$ that is larger than the actual velocity. This is due to the addition along the LOS of various velocity components, either from other strands and/or from the surrounding corona of each strand, which moves $90^{\circ}$ out-of-phase to the strand. In $x \sim 0.3\Mm$ there are two strands in the same LOS (marked with a green arrow). In this region, there is a clear interference pattern associated with the integration of the Doppler signal along the LOS. 

For the case of 30 strands the situation is more complex (Fig. \ref{fig:synthetic}(b)). In the central region, there are several strands for each LOS. The Doppler pattern is very complex in the central region. The reason is that the integration of the Doppler signal along the LOS can produce constructive or destructive interference patterns. For example, in $x=-1.6\Mm$ (marked with a green arrow) there is a minimum of the Doppler signal but a large line broadening. This indicates opposite motions of the plasma in this LOS that produce a cancellation of the Doppler signal and a line broadening. It is important to note that the fundamental vibrations considered here do not perturb much the temperature. Hence, the line broadening mostly corresponds to non-thermal line broadening and therefore gives an estimate of the variance of the Doppler velocity along the LOS. In contrast, at $x=0.4\Mm$ (light blue arrow) there is an enhancement of the Doppler signal with a moderate broadening signal suggesting a constructive interference along the LOS. Around $x\sim -0.7\Mm$  (yellow arrow) there are large Doppler velocity and line broadening signals indicating that the Doppler signal is showing the motion of an individual strand.  

Figure \ref{fig:synthetic}(c) shows the most complex situation with 100 strands. Both Doppler velocity and line broadening time-distance diagrams show very complex patterns. The green and blue arrows show examples where the integration produces destructive or constructive interference respectively. Due to the complexity of the patterns it is difficult to distinguish regions with significant Doppler signal and a large broadening associated with the collective oscillations of the strands. The two-headed yellow arrow shows the region where a collective oscillation starts and ends. However, in most cases, it is almost impossible to isolate the individual strand motions from their surroundings due to the complexity of the diagrams. As in the situation in the case marked with the yellow arrow, they appear surrounded by constructive and destructive patterns.

In order to distinguish the collective oscillations with respect to a non-interacting system, we have considered the same 100 strand distribution but with each strand oscillating individually with a kink mode. The initial phase of the oscillations and the direction of the motion have been randomly generated between 0 and $2\pi$. The oscillation amplitude of each strand is fixed to $10\kms$. The strand densities have been randomly generated in order to have a kink oscillation period between 4.637 and 5.338 min as the range of periods of the collective modes. Figure \ref{fig:synthetic}(d) shows the Doppler velocity and line broadening time-distance diagrams. The Doppler velocity pattern also shows a complex structure of cancellation and superposition of the signals. It is thus difficult to distinguish it from the interacting strands case (Fig. \ref{fig:synthetic}(c)). In contrast, the line-broadening diagram is clearly different from the interacting strands case. The line-broadening appears more uniform and individual strands seem more easily distinguishable. The reason is that the individual strands oscillate always with the same velocity amplitude. Regions with large broadening are produced when two strands oscillate in opposite directions. Conversely, regions with a  reduction of the line broadening coincide with regions with an increase of the Doppler velocities due to constructive interference. In this situation, the complexity of the diagrams is associated with the interference pattern of the motions of the strands but not with the complexity of the motions of the strands.

\begin{figure*}[!ht]
\begin{center}
\includegraphics[width=0.999\textwidth]{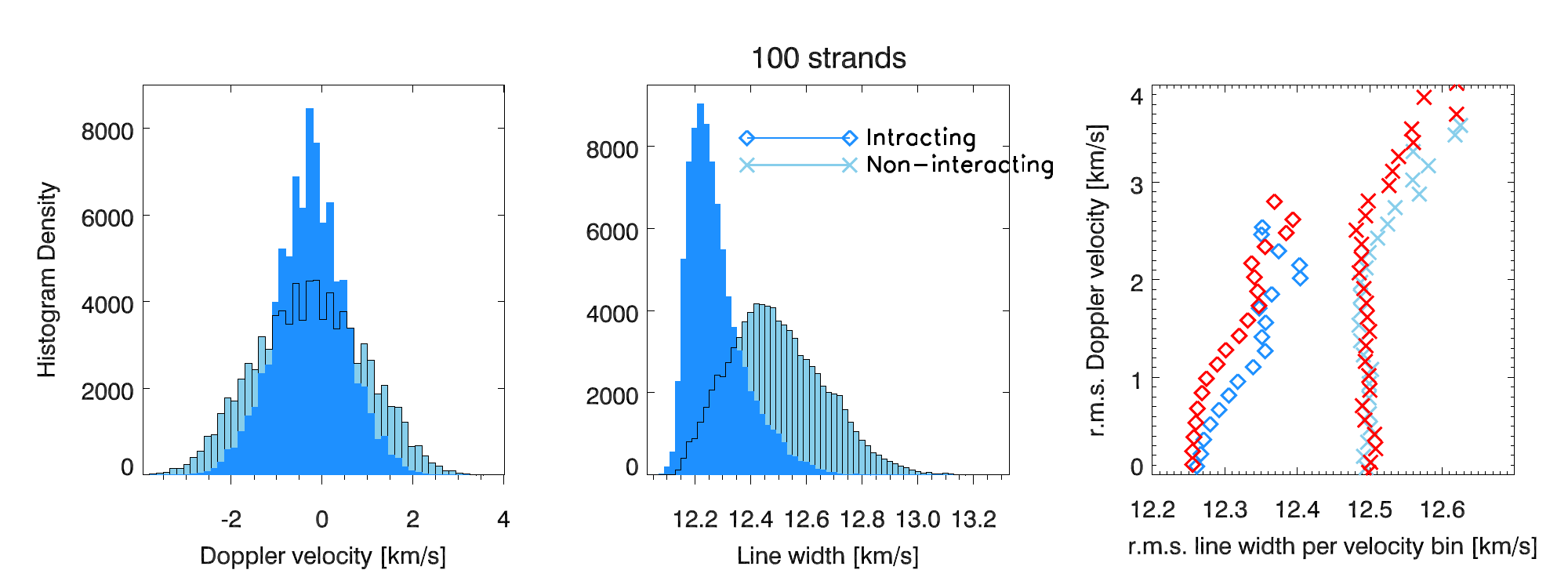}
\end{center}
\caption{Plot of (a) the Doppler velocity and (b) the line width distributions for the 100 strand case with interactions (``interacting'') and without interactions (``non-interacting''). In (c) we plot the r.m.s. value in line width for each bin of the velocity histogram against the r.m.s. value for the same pixels. Blue and red symbols in this panel correspond, respectively, to positive and negative Doppler velocities.} \label{fig:histograms}
\end{figure*}

Figure \ref{fig:histograms}(a) shows the histogram of the Doppler velocities in both cases of 100 strands with interaction (`vibrations') or without interaction (`random'). 
In the non-interacting tubes case, both the Doppler velocity and line broadening distributions are broader. In the interacting tubes case, the motions of the strands are more complex, with rotations of the direction of the kink oscillations, damping and amplifications producing narrower distributions. Similarly, the peak of the line-broadening distribution for the interacting tubes case has a smaller value, indicating less destructive interference. In panel \ref{fig:histograms}(c) we have plotted the r.m.s value of the Doppler velocities as a function of the line broadening velocities. Both situations are clearly different. For the interacting strands, there is a clear positive correlation between both magnitudes for Doppler velocities smaller than $1.2\kms$. This is due to the existence of regions of amplified Doppler velocity and line-broadening associated with the individual motions of the strands that we have discussed previously. While for the non-interacting case, any LOS will have, on average, the same variation of Doppler velocities, thereby leading to a roughly equal line-width and no correlation. For Doppler velocities larger than $1.2\kms$ the correlation is unclear and/or statistically not significant since the number of pixels in the histogram bins at these high velocity values is small. The correlation between the line broadening and the Doppler velocities is an important result that may allow to distinguish between random individual oscillations and collective vibrations of the corona.

\section{Discussion and Conclusions}\label{sec:conclusions}
In this work, we study the collective oscillations of the AR corona under the hypothesis that it is composed of myriads of strands. We have found that in the collective modes, not all the strands participate in the motion. The modes are formed by \emph{clusters} of tubes oscillating in different ways. Additionally, the modes with the lowest and the highest periods form a \emph{chain} of strands. The periods distribute over a wide band of values. The width of the band increases with the number of strands but rapidly reaches an approximately constant value. Thus, for a sufficiently large number of strands the period range is independent of the size of the system. We have found an approximate expression for the minimum and maximum periods of the band. Our findings indicate that the frequency band associated with the fine structure of the corona depends on the strand configuration and the minimum distance between strands. The chain of strands associated to the highest and lowest periods are formed by the strands with the shortest distance between them. The size of this cluster is independent of the size of the system. However, for the rest of modes the cluster sizes increase with the size of the system. This indicates that the motion of one strand is influenced by the motions of distant strands. In this sense, in a finite volume of the corona, the oscillations of the strands inside the region are coupled to the strands outside of the region.

We have produced synthetic images of Doppler velocities and line broadening in order to find observational signatures of these fundamental vibrations. We consider the strands in a finite volume of the corona and the synthetic images are time-distance diagrams in a slit perpendicular to the axis of the strands. The strands of the region vibrate with a linear combination of collective normal modes. Both Doppler velocity and line width diagrams show a complex pattern. Part of this complexity is associated with the cancellation and superposition of the velocities along the LOS. However, the collective motions of the strands introduce additional complexity to the diagrams in terms of additional ordered motions (or less randomness in the system). We have generated a situation without interaction (i.e. pure random motions) between strands and we have found a clear difference. In the non-interacting strands situation, the line broadening is not correlated with the Doppler velocity. In contrast, in the interacting tubes case, the line broadening shows a positive correlation with Doppler velocities. {We interpret this as a signature of the collective nature of the strand oscillations however it is a weak confirmation of the model because other models can also produce a positive correlation. Computational limitations restrict our models to systems with a maximum of 100 strands. Increasing the number of strands will result in collective effects affecting more strands. It will increase the non-uniformity of the collective signatures. In addition, the consideration of the background and foreground corona can reduce or amplify these collective signatures. Observational signatures from improved models will display more complex line broadening and Doppler velocity patterns. This work motivates future research in this direction by considering a much larger strands system.}  
Although we have focused in this work on the assumption of a corona, which is composed of strands, the same results apply to a multitude of flux tubes distributed in a similar fashion. We have found that due to the nature of the solar corona composed of elemental strands and/or flux tubes, it harbours a kind of collective vibrations that we have called fundamental vibrations. These vibrations cannot exist in a uniform corona or without coupling between the motion of strands.

This work is a first step to further consider fast MHD wave propagation in the corona and the influence of the fine structure on it. The waves are resonantly scattered in the non-uniform corona. The frequencies of the waves that are scattered are the frequencies of the fundamental vibrations of the corona. The existence of the fine structure can affect the propagation of the energy and the coronal energy budget as has already been pointed out by \citet{moortel2012,VanDoorsselaere:2014it}. The propagation of waves in a structured corona will be considered in a future study. 
{ Should results on the collective nature of periodic standing waves remain valid for propagating waves,  this would reinforce the conclusion by \citet{mcintosh2012} regarding the origin of the positive correlation between Doppler velocities and line broadening being the kink wave nature of propagating waves.}

\begin{acknowledgements}
M. Luna acknowledges the support by the Spanish Ministry of Economy and Competitiveness (MINECO) through projects AYA2014-55078-P and under the 2015 Severo Ochoa Program MINECO SEV-2015-0548. R.O. acknowledges support from the Spanish Ministry of Economy and Competitiveness (MINECO) and FEDER funds through project AYA2017-85465-P. P.A. acknowledges funding from his STFC Ernest Rutherford Fellowship (No. ST/R004285/1). M.L., R.O. and P.A. acknowledge support from the International Space Science Institute (ISSI), Bern, Switzerland to the International Team 401 `Observed Multi-Scale Variability of Coronal Loops as a Probe of Coronal Heating' (P.I. Clara Froment and Patrick Antolin). I.A. acknowledges financial support from the Spanish Ministerio de Ciencia, Innovaci\'on y Universidades through project PGC2018-102108-B-I00 and FEDER funds.
\end{acknowledgements}


\end{document}